\documentclass[twocolumn,astrosymb]{aastex702}

\usepackage{color}
\usepackage{comment}

\newcommand{\cmm}{\ifmmode{\rm cm^{-2}}\else{$\rm cm^{-2}$}\fi}
\newcommand{\hi}{\ifmmode{\rm HI}\else{H\/{\sc i}}\fi} 
\newcommand{\oh}{\ifmmode{\rm OH}\else{O\/{\sc H}}\fi} 
\newcommand{\glon}{\ifmmode{\ell}\else{$\ell$}\fi} 
\newcommand{\glat}{\ifmmode{b}\else{$b$}\fi}
\newcommand{\vlsr}{\ifmmode{V_\mathrm{LSR}}\else{$V_\mathrm{LSR}$}\fi}
\newcommand{\vwind}{\ifmmode{V_\mathrm{w}}\else{$V_\mathrm{w}$}\fi} 
\newcommand{\dg}{\ifmmode{^\circ}\else{$^\circ$}\fi} 
\newcommand {\kms}{\ifmmode{\rm km \, s^{-1}}\else{$\rm km \, s^{-1}$}\fi}

\newcommand{\nhi}{\ifmmode{N_{\rm HI}}\else{$N_{\rm HI}$}\fi}
\newcommand{\noh}{\ifmmode{\rm N_{OH}}\else{$\rm N_{OH}$}\fi}
\newcommand{\nh}{\ifmmode{\rm N_{H}}\else{$\rm N_{H}$}\fi}
\newcommand{\vout} {\ifmmode{V_\mathrm{out}}\else{$V_\mathrm{out}$}\fi} 
\newcommand{\Vr}{\ifmmode{V_{r}}\else{$V_{r}$}\fi}
\newcommand{\vrmin}{\ifmmode{V_{\rm r,\,min}}\else{$V_{\rm r,\,min}$}\fi}
\newcommand{\TL}{\ifmmode{T_{\rm L}}\else{$T_{\rm L}$}\fi}
\newcommand{\dsun}{\ifmmode{d_{\rm Sun}}\else{$d_{\rm Sun}$}\fi}
\newcommand{\cosbeta}{\ifmmode{\rm cos(\beta)}\else{${\rm cos(\beta)}$}\fi}

\begin{document}
\pagenumbering{arabic}
\shorttitle{Fermi Bubble \hi\ Clouds}
\title{High Velocity Neutral Gas in the Fermi Bubbles: New Kinematic Limits and Spatial Structure}

\author[0000-0002-6050-2008]{Felix J. Lockman}
\affiliation{Green Bank Observatory, National Radio Astronomy Observatory, Green Bank, WV 24944, USA}
\email[show]{jlockman@nrao.edu}

\author[0000-0003-4019-0673]{Enrico M. Di Teodoro}
\affiliation{Universit\`{a} di Firenze, Dipartimento di Fisica e Astronomia, Largo Enrico Fermi 2, 50125 - Firenze, Italy}
\affiliation{INAF - Osservatorio Astrofisico di Arcetri, Largo Enrico Fermi 5, 50125 - Firenze, Italy}
\email{enrico.diteodoro@unifi.it}

\author[0000-0002-6050-2008]{Savannah Cary}
\affiliation{Green Bank Observatory, National Radio Astronomy Observatory, Green Bank, WV 24944, USA}
\affiliation{Department of Astronomy, University of California Berkeley}
\email{scary@berkeley.edu}

\author[0000-0003-0724-4115]{Andrew J. Fox}
\affiliation{AURA for ESA, Space Telescope Science Institute, 3700 San Martin Drive, Baltimore, MD 21218, USA}
\email{afox@stsci.edu}

\begin{abstract}
We have detected hundreds of neutral clouds entrained in the Milky Way's nuclear wind using \hi\ data from new surveys made with the Green Bank Telescope that cover about 500 sq-degrees around the Galactic center (GC).  
Galactic winds are common throughout the Universe, and these data at $9.1'$ angular resolution (22 pc at the GC) provide the most detailed analysis of the vertical profile of a neutral nuclear wind in any galaxy. 
A set of 228 of these Fermi Bubble clouds with the largest values of $|\vlsr|$ has been analyzed to examine the distribution and kinematics of the outflowing gas. 
The clouds span $-335 \ \kms\ \leq \vlsr \leq +438$ \kms, the largest positive LSR velocities ever reported for neutral \hi\ associated with the Milky Way disk. 
The highest velocities are found furthest from the GC, suggesting that clouds are accelerated from a low velocity near the nucleus to at least 500 \kms\ at a radial distance of $\lesssim 4$ kpc. 
Clouds appear disrupted as they are accelerated: their line brightness and \nhi\ decreases steadily with distance from the GC, and the population becomes more uniform. 
There is an abrupt  cutoff in the neutral clouds at a vertical distance of $\approx2$ kpc from the Galactic plane. 
Kinematic models of an outflowing cloud population that fills the FB volume are used to identify structure  in the gas. 
The kinematics of the highest velocity, highest latitude clouds imply a past azimuthal asymmetry in the outflow. 
\end{abstract}


\section{Introduction} 
\label{sec:introduction}
The Milky Way has an extended central region defined by vertical outflow of energetic particles from its nucleus.
This region is visible across the electromagnetic spectrum, and is commonly called the Fermi Bubbles (FBs) \citep[e.g.,][]{Bland-Hawthorn_Cohen03, Keeney2006,Su+10, Carretti2013}.
The FBs are of particular interest as nuclear outflows are a common phenomenon and can have a significant effect on the structure and evolution of their host galaxies \citep{Veilleux+05,Veilleux2020,Heckman2017,Tumlinson2017}.
In the Milky Way, the FBs have interacted with the Galactic interstellar medium over an area many square-kpc in size, covering many hundreds of square-degrees around the Galactic center \citep{Lockman84, Lockman_McClureGriffiths16,Sofue2021,SuZhang2022}.
Even more extended structures are visible in X-ray emission suggesting that the phenomena may have resulted from multiple events \citep{Predehl2020}. 
The contents of the outflowing material are manifest not only in continuum emission but also in absorption lines of ionized gas \citep{Keeney2006,Fox+15,Savage+17}, and emission from neutral atomic and molecular clouds \citep{McClure-Griffiths+13,DiTeodoro18,Lockman2020,DiTeodoro2020,DiTeodoro2026a}.
These spectroscopic studies offer the promise of determining the kinematics, mass outflow, and energetics of the phenomenon by measuring the properties of gas entrained in the hot nuclear wind. 
For a recent review of the FBs and related phenomena, we refer to \citet{Sarkar2024}.

There are cloud-like structures seen in \hi\ at low Galactic latitude throughout the inner Galaxy \citep[e.g.,][]{Lockman2002,Stil2006,Ford2010}.
These disk clouds, however, have velocities entirely consistent with Galactic rotation to within a few tens of \kms.
The \hi\ cloud population toward the FBs, first identified by \citet{McClure-Griffiths+13}, deviates from allowed motions by more than 100 \kms\ at both positive and negative Local Standard of Rest (LSR) velocities,  even when potential streaming motions induced by a bar are considered \citep[e.g.,][]{Binney1991,Weiner1999,Sormani2015}.

In this paper we describe a new Green Bank Telescope (GBT) survey of 
\hi\ emission clouds associated with the FBs with detections at higher velocities and larger distances from the Galactic center (GC) than were known previously.
A set of clouds having LSR velocities $|\vlsr| \gtrsim 130$ \kms\ is analyzed for information on the kinematics of the outflow.
Because the survey covers a very large volume of the FBs, the interpretation of the results must take into account the exact sky coverage, sensitivity issues in cloud detectability, and cloud evolution within the FBs -- factors that were less critical in previous, smaller surveys.
This attention to detail allows us to understand several features in the data that arise from observational issues and not real structure in the neutral outflow.

A key assumption in this paper is that the observed \hi\ clouds originated near the GC and are entrained in a hot fast wind that has accelerated the clouds to their observed velocities, which exceed anything found elsewhere in the Milky Way disk.
We also assume that the clouds are confined within the boundaries of the FBs,  although we cannot directly connect the process that has accelerated the neutral clouds to the process that created the FBs.

The paper is structured as follows.
In Section~\ref{sec:observations} we describe the observations and  data  reduction, and in Section~\ref{sec:cloudID} the 
selection of the sample of \hi\ clouds that will be analyzed.
Section~\ref{sec:cloud_properties} gives the properties of the clouds  derived directly from the observations, and compares the data for the Northern FB with that of the Southern FB, concluding that their kinematics are  similar.
Section~\ref{sec:geometry} describes the  geometry of the FBs and how any  outflow velocity projects onto the observable \vlsr.
In Section~\ref{sec:Vout_limits} we  derive limits on the intrinsic outflow velocity of the \hi\ clouds and find that the data require an outflow accelerating from low velocities close to the GC to $\sim500$ \kms\ at a distance $\sim 4$  kpc from the GC.
Using a first-order model for the outflow velocity, we derive several aspects of the \hi\ cloud population in Section~\ref{sec:variation};  this  reveals a decline in the brightness of the 21cm line and \nhi\ of clouds with increasing distance from the GC, strongly suggesting cloud evolution from compact to diffuse.
Section~\ref{sec:detectability} considers observational effects on cloud detectability and Section~\ref{sec:models} describes our simulations of the cloud population.
In Section~\ref{sec:comparison_with_obs} several simulations are compared with the data giving information on  structure within the neutral outflow. 
In Section \ref{sec:long-vel} we describe asymmetries in the longitude-velocity distribution of the clouds with implications for the history of the outflow.
We conclude in Section \ref{sec:summary} with a summary and discussion of how these findings relate to our understanding of the FB outflow, and list several issues that remain unresolved.
Throughout the paper, we assume a distance to the GC of 8.275 kpc \citep{GRAVITY2021}. 
At this distance the $9.1'$ angular resolution of the GBT corresponds  to 22 pc.

\section{Observations and Data Reduction}
\label{sec:observations}

The 21cm \hi\ data presented here are from new surveys of \hi\ emission in the FBs made with the GBT \citep[][]{Prestage2009} between 2020 and 2022, and data from previous GBT surveys  presented in \citet{DiTeodoro18} and \citet{Lockman2020}.
The data set analyzed here results from a complete re-reduction of all existing GBT \hi\ data 
 as well as inclusion of the new data which extends the observed area in longitude and latitude, and additional observations of previously measured regions at greater sensitivity.

In the 21cm line of \hi\ the GBT has an intrinsic angular resolution of $9.1\arcmin$. 
The FBs were  mapped in $2^{\circ} \times 2^{\circ}$ tiles where  data were taken using ``on-the-fly"  sampling every 1.875\arcmin\   in longitude over 2-degree long strips  spaced 3.75\arcmin\ in latitude.
The GBT L-band receiver was used for all observations.
The new spectra were acquired with the VEGAS spectrometer \citep{Prestage2015} and combined with  the older data some of which was taken with the GBT spectrometer \cite[see][]{DiTeodoro18}.
In all cases in-band frequency switching was used and the intrinsic spectral resolution of $<0.1$ \kms\ was smoothed to a uniform 3.02 \kms\ per spectral channel.

\subsection{Calibration and data reduction}
\label{sec:calibration}

Spectra were calibrated and corrected for stray radiation as described in \citet{Boothroyd2011}, although for the velocities of the \hi\ clouds discussed here, stray radiation is generally negligible.
There was occasional narrow-band radio frequency interference (RFI) that was removed by interpolating over the affected channels.
While evidence of some residual RFI can be seen in the data, in no case does it compromise measurements of individual \hi\ clouds.
Spectra were smoothed to a channel spacing of 3.02 \kms, and were gridded into a data cube with a pixel size of $200''$ using the \texttt{gbtgridder} software.
A polynomial was then fit to emission-free regions of the spectra pixel by pixel using an iterative process to mask channels with significant \hi\ emission.
The final data cube covers a velocity range of $-585 \leq \vlsr \leq +584$ \kms. 

\begin{figure}
\includegraphics[width=0.47\textwidth]{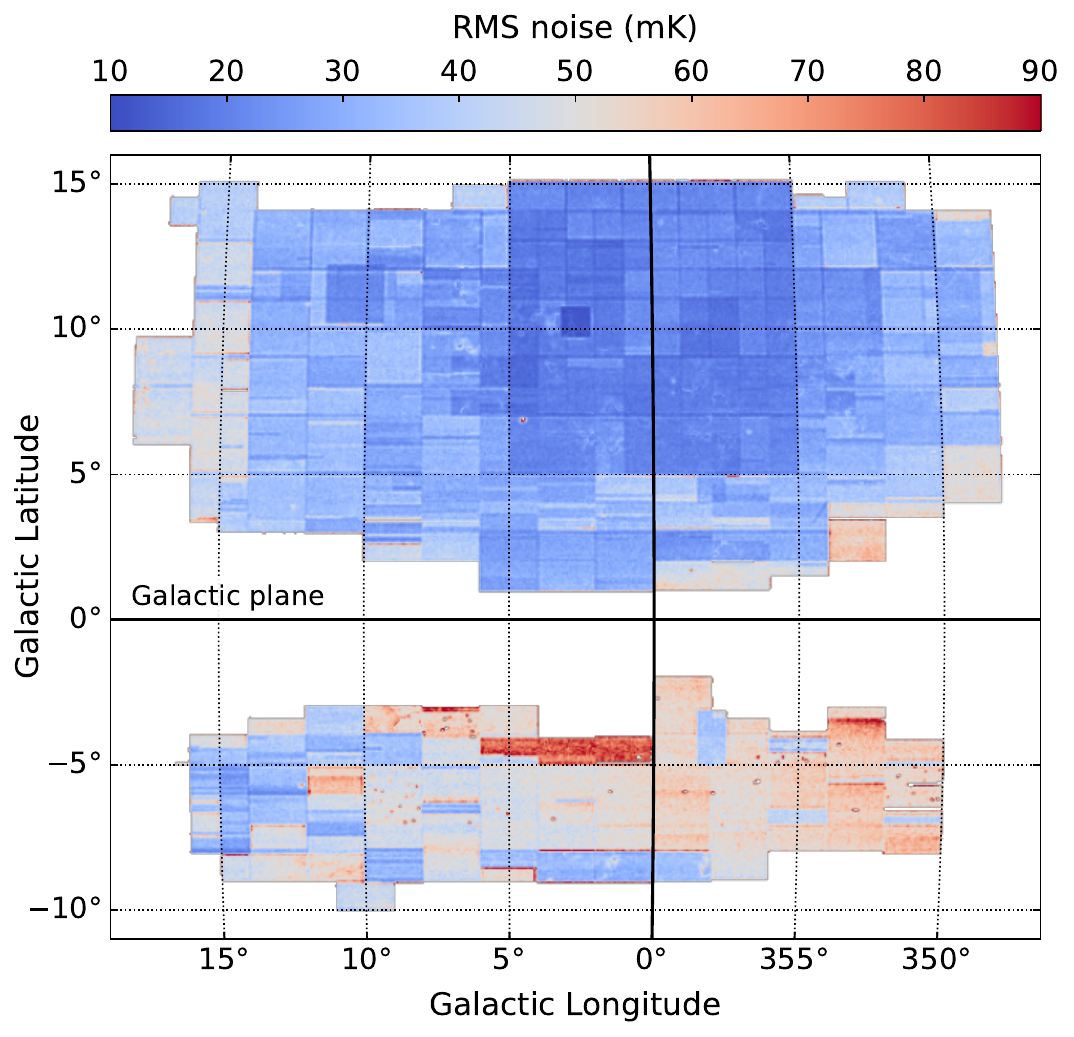}
\caption{
The rms noise in a spectral channel over the region of the GBT survey.
This map is for the final data cube at 3.02 \kms\ velocity resolution.
}
\label{fig:noisemap}
\end{figure}

Figure \ref{fig:noisemap} shows the total sky coverage of the survey color coded by the root-mean-square (rms) noise.
A 100 sq-deg region between $-5^{\circ} \leq \ell \leq +5^{\circ}$ and $+5^{\circ} \leq b \leq +15^{\circ}$ was observed extensively to a 3.0 \kms\ channel  rms noise of 20.5 mK, and the 
 $2\dg \times 2\dg$ area around the Active Galactic Nucleus (AGN) PDS 456 at $(\ell,b) = (10.4\dg,+11.2\dg$) \citep[see][]{Fox+15} was observed to a rms noise of 22 mK.
We note that while the spectral noise varies across the survey area, in each spectrum the noise is identical at positive and negative velocities, and thus the detectability of clouds is  independent of their \vlsr.

\section{Identification of Fermi Bubble clouds}
\label{sec:cloudID}

We examined the final data cubes for \hi\ emission that is  isolated in position and velocity from \hi\ in the Milky Way disk.
For the study presented here we considered only emission at $|\vlsr| \gtrsim 130$ \kms\ because it is these clouds that set the most strict limits on the kinematics of the population.
The \hi\ clouds in the FB wind often have a complex structure with several components \citep{Noon2023} which can make the process of counting clouds fraught with ambiguity.
This is particularly true for relatively low-velocity objects, whose emission may overlap spatially or spectrally with foreground emission from the Milky Way disk.
For $|\vlsr| \gtrsim 130$ \kms, however, there is little blending of emission features and it is generally straightforward to identify individual clouds that can be considered as independent entities. 

Cloud properties were determined using the Cube Analysis and Rendering Tool for Astronomy \citep[CARTA,][]{Wang+2026}.  
Clouds were identified in channel maps and spectra were derived from averages  over a $10\arcmin \times 10\arcmin$ area around the emission peak.  
Using CARTA, a Gaussian function was fit to the spectra to derive the peak line brightness temperature (\TL), line Full Width at Half Maximum (FWHM), LSR velocity (\vlsr), and their associated errors.
The  \hi\ column density, \nhi, was calculated from the parameters derived from the Gaussian fit: $\nhi = 1.82\times 10^{18} \times \TL\ \times {\rm FWHM} \times 1.065 \ \cmm$, under the optically thin assumption that is appropriate for clouds of such low \nhi\ \citep{DickeyLockman1990}. 
Errors on \nhi\ were propagated from the Gaussian fit.
Occasionally the \hi\ spectra showed multiple components arising from  confusion of several clouds at similar positions and velocities.
If the cloud spectral components could not be determined unambiguously, the cloud was dropped from the sample, though 
this occurred in only a few percent of the cases.
From the Gaussian fits a few clouds were found to have $|\vlsr|$ slightly less than 130 \kms\  but were still included in the data set, so the velocity cutoff is not strict.

\begin{figure}
\includegraphics[width=0.47\textwidth]{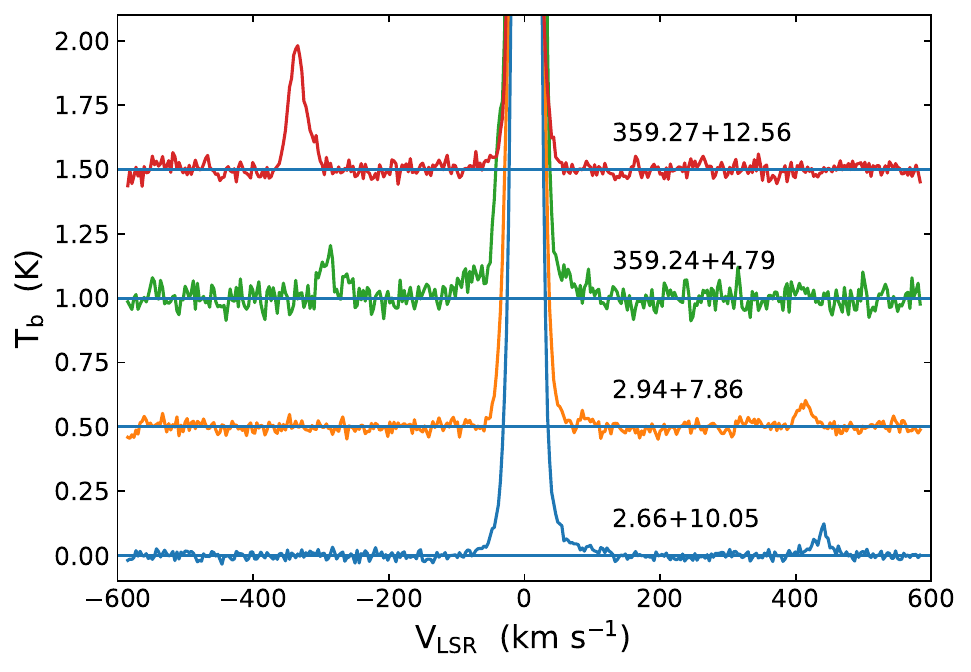}
\caption{Selected spectra from the GBT survey showing the line strength of some of the highest-\vlsr\ clouds at the location of their emission peak.
}
\label{fig:clouds_spectra}
\end{figure}

Table \ref{tab:cloud_table} shows a sample of the cloud catalog and 
a machine-readable table containing all derived quantities for individual clouds is provided as supplementary online material.
Figure~\ref{fig:clouds_spectra} shows spectra for some of the clouds in the survey from the pixel at the location of the cloud's  brightest emission.
In all cases, the high-velocity \hi\ emission of clouds is clearly detected and well separated from the dominant emission at $\vlsr\simeq0$, which arises from the local interstellar medium in the Solar neighborhood.
   
\begin{deluxetable*}{cccccccccc}
\label{tab:cloud_table}
\tablecolumns{10}
\tablewidth{0pt}
\tablecaption{Properties of Individual GBT clouds  \tablenotemark{a}
\label{:tab:cloud_measurements}}
\tablehead{
\colhead{ID}&
\colhead{$\ell$}&
\colhead{$b$}&
\colhead{$T_{\rm L}$}&
\colhead{${\rm FWHM}$} &
\colhead{${\rm VLSR}$}&
\colhead{$\nhi$} \\
\colhead{}&
\colhead{[\dg]} &
\colhead{[\dg]} &
\colhead{[K]}&
\colhead{(\kms)} &
\colhead{(\kms)}&
\colhead{${\rm (10^{19} \cmm)}$}        }
\startdata
1 & 359.27 & 12.56 & $0.342\pm0.008$ & $27.3\pm0.7$ & $-335.2\pm0.3$ & $10.0\pm0.2$\\
2 & 359.24 & 4.79 & $0.135\pm0.009$ &  $22.9\pm1.8$ &  $-291.8\pm0.8$ & $3.3 \pm0.2$\\ 
3 & 356.75 & 13.23 & 0.047$\pm$0.030 & 26.3$\pm$19.6 & $-285.2\pm19.6$ & $1.3\pm0.8$ \\
4 & 1.31 & 6.00 & 0.067$\pm$0.006 & 38.4$\pm$3.8 & $-284.2\pm3.8$ & $2.7\pm0.2$ \\
5 & 355.32 & 5.34 & 0.097$\pm$0.010 & 19.1$\pm$2.2 & $-274.9\pm2.2$ & $2.0\pm0.2$ \\
6 & 0.94 & 6.34 & 0.134$\pm$0.004 & 34.4$\pm$1.3 & $-264.5\pm1.3$ & $4.9\pm0.2$ \\
225 & 2.94 & 7.86 & 0.065$\pm$0.007 & 23.1$\pm$2.7 & $+411.7\pm2.7$ & $1.6\pm0.2$ \\
226 & 2.68 & 10.21 & 0.048$\pm$0.006 & 23.2$\pm$3.5 & $+426.5\pm3.5$ & $1.2\pm0.1$ \\
227 & 2.89 & 9.89 & 0.044$\pm$0.006 & 33.1$\pm$5.1 & $+429.2\pm5.1$ & $1.5\pm0.2$ \\
228 & 2.66 & 10.05 & 0.049$\pm$0.011 & 19.7$\pm$5.1 & $+437.9\pm5.1$ & $1.0\pm0.2$ \\
\enddata
\tablenotetext{a}{Uncertainties are $1\sigma$ from the Gaussian fit.}
\end{deluxetable*}

The survey area can be roughly divided into three regions of differing noise level (Figure~\ref{fig:noisemap}).
For $b > 0\dg$ the rms noise in a 3 \kms\ channel has an average value of $\langle\sigma\rangle = 30$ mK, except within the ``deep" area where $\langle\sigma\rangle =  20.5$ mK, while at $b < 0\dg$ the  $\langle\sigma\rangle =  52$  mK.
The corresponding limits on \nhi\ can be derived assuming detection at the  $3\sigma$ limit on \TL, and a 24 \kms\ FWHM, which is the median value for clouds in the survey.  
For the three regions this gives $3\sigma$ limits on \nhi\ of $\leq4.2 \times 10^{18} \, \cmm$, $\leq2.9 \times 10^{18} \, \cmm$, and $\leq7.3 \times 10^{18} \, \cmm$.
The observed distributions of \TL\ and \nhi\ described in the next section are consistent with these limits on survey completeness.

\begin{figure}
\includegraphics[width=0.47\textwidth]{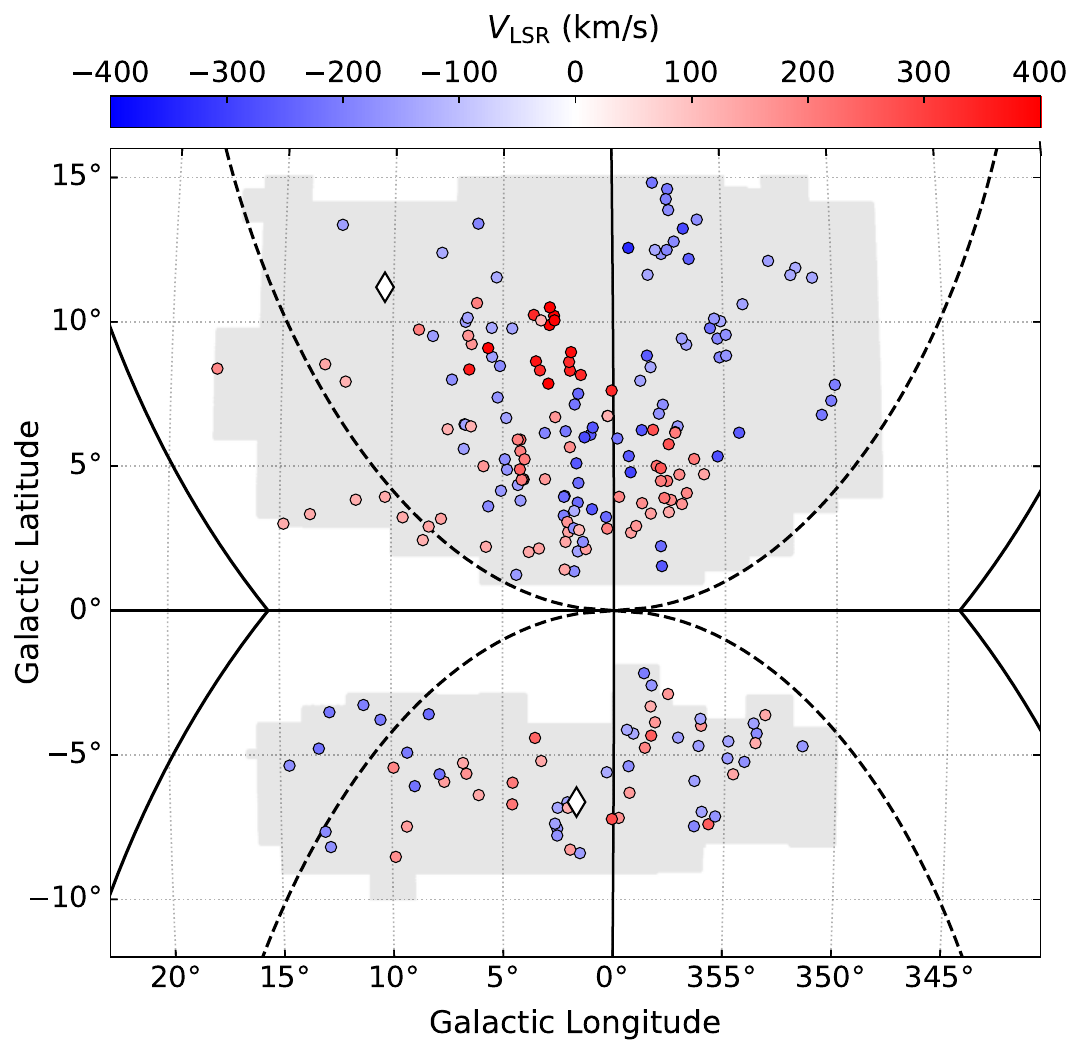}
\caption{
The location of clouds in the sample colored by their \vlsr. 
The survey boundaries are shown by the shaded areas.
The locus of the FBs is taken from the \citet{MillerBregman16} model as viewed from the Sun, where solid curves mark the outer shock boundary and dashed curves the inner filled bubble.
Open black diamonds denote the location of two UV-bright background objects toward which absorption spectra of ionized species in the  FBs have been measured.
}
\label{fig:clouds_l-b-VLSR}
\end{figure}

The catalog contains 228 clouds.
Figure \ref{fig:clouds_l-b-VLSR} shows the location of the \hi\  clouds in our catalog colored by their \vlsr. 
The shaded areas  show the  boundaries of the GBT survey (Fig.~\ref{fig:noisemap}). 
The solid and dashed black curves trace the projected edges of the \citet[][hereafter MB16]{MillerBregman16} model for the thermal hot gas within the FBs. 
This model, which is based on X-ray data of the \ion{O}{7} and \ion{O}{8} emission lines, contains two components: an outer shocked ``shell" of compressed material (solid line) and an inner plasma-filled ``bubble", co-spatial with the $\gamma$-ray FBs (dashed line).
We detect clouds outside of the inner filled shell and thus, for the remainder of this paper, we will adopt the MB16 outer shock locus as the outer boundary of the  FBs.

%%%%%%%%%%%%%%%%%%%%%%%%%%%%%%%%%%%%%%%%%%%%%%%%%%%
%%%%%%%%%%%%%%%%%%%%%%%%%%%%%%%%%%%%%%%%%%%%%

\subsection{Related results from UV spectroscopy}
\label{sec:UV}

There is information on ionized FB gas in the area of our survey from ultra-violet (UV) absorption spectra taken with the Hubble Space Telescope (HST) towards two background objects: the AGN PDS 456 at $(\ell,b) = (10.4\dg,+11.2\dg)$ \citep{Fox+15} and the distant star LS 4825 at $(\ell,b) = (1.76\dg,-6.63\dg)$ \citep{Savage+17}.
These absorption lines arise from species such as \ion{C}{4}, \ion{Si}{4} and \ion{N}{5}.
The UV spectra reveal the presence of ionized gas whose physical properties are very different from the \hi\ clouds, but this gas is certainly entrained in the FB wind and its kinematics are an important addition to our study. 
The location of the AGN PDS 456 and of the star LS 4825 sightlines are shown as open diamonds in Figure~\ref{fig:clouds_l-b-VLSR}.

The area around PDS 456 was singled out for special consideration in the GBT \hi\ observations because  the absorption lines  in this direction extend over  $-235 \leq  \vlsr \leq +250$ \kms, but no \hi\ emission at the corresponding high velocities was detected to a $3\sigma$ limit  of $3.1 \times 10^{18} \ \cmm$ for a FWHM of 24 \kms.
Toward the distant star LS 4825 there is absorption from ions covering $-250 \leq \vlsr \leq +150$ \kms, but the \hi\ emission is restricted to $-105 \leq \vlsr \leq +98$ \kms.
The $3\sigma$ limit on higher $|\vlsr|$  \hi\ for LS 4825 is $7.1 \times 10^{18} \ \cmm$.

Several FB sightlines outside the current GBT survey area show \hi\ emission and UV metal absorption from the same clouds, with matching velocity centroids.
This includes the clouds seen in the area toward the AGN 1H1613-097 at $b = +28.5\dg$ \citep{Bordoloi+17, Bordoloi2025}, and several FB clouds at $|b| > 18\dg$ 
with metallicity measurements \citep{Ashley2022}.
Thus there can be multi-phase structure in the FB clouds, with regions of neutral and ionized gas lying close together.
The physical relationship between the high-latitude (mostly UV-traced) and the low-latitude (mostly \hi-traced) clouds remains unclear.

%%%%%%%%%%%%%%%%%%%%%%%%%%%%%%%%%%%%%%%%%%%%%%%%%%%%%%% Properties
\section{Properties of the detected \hi\ Clouds}
\label{sec:cloud_properties}

In this section we describe the main features of the GBT cloud sample.

\begin{figure*}
\centering
\includegraphics[width=0.95\linewidth]{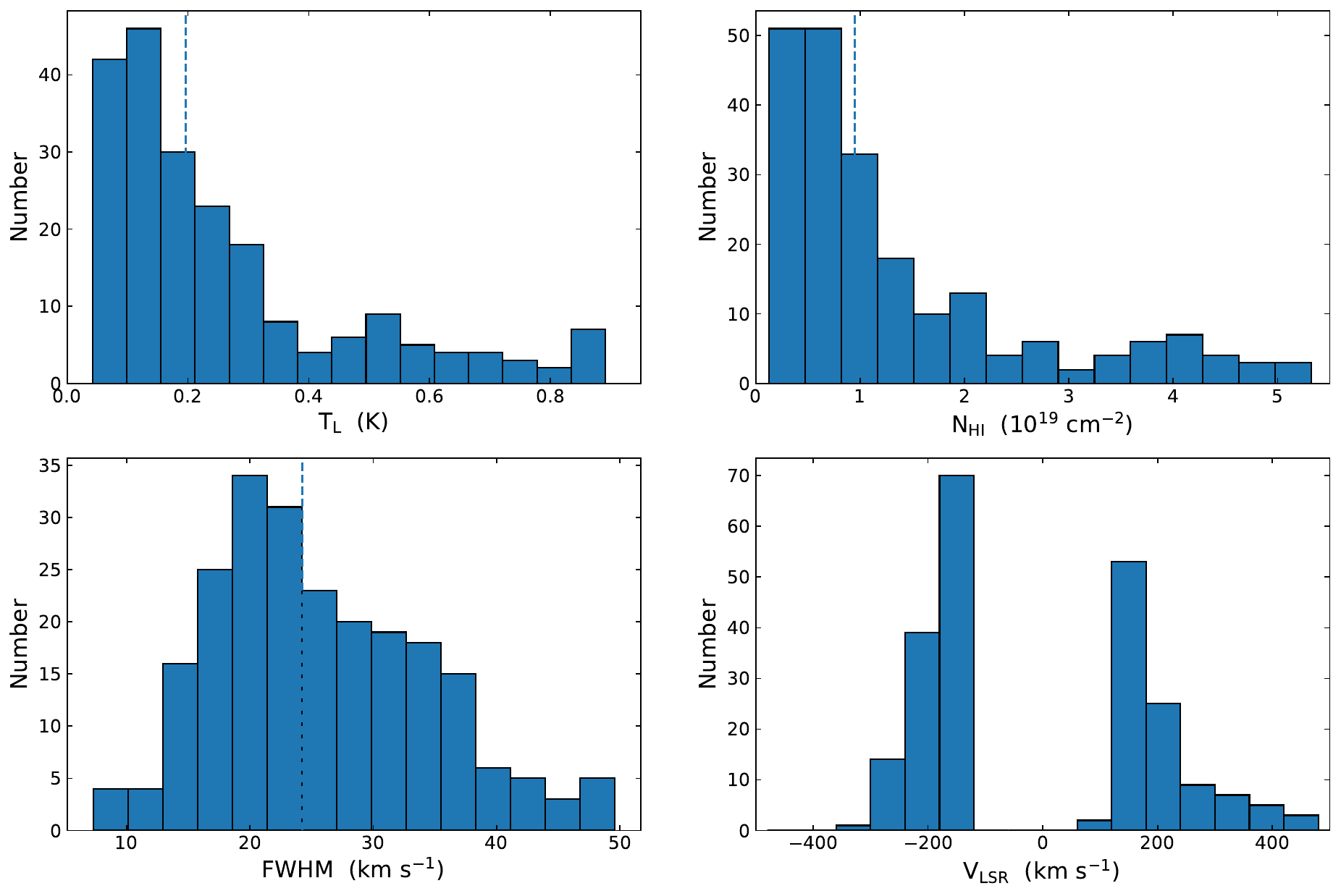}
\caption{
{\it Upper Left panel:} 
The peak \TL\ of \hi\ lines in the  sample, derived from a Gaussian fit.
The median, marked with dashed line, is 0.20 K.
This figure shows the clouds with $\TL \leq 1$ K ($92\%$ of the sample).
Another $6.5\%$ have $1 \leq \TL\ \leq 2$ K.
{\it Upper Right panel:}
The \hi\ column density \nhi\ averaged over a $10\arcmin \times 10\arcmin$ area at  the emission peak of each cloud.
This shows $94\%$ of the clouds; the others have $\nhi > 5.5 \times 10^{19}$ \cmm.
The median \nhi, marked with the dashed line, is $0.95 \times 10^{19} \ \cmm$.
{\it Lower Left panel:}
The \hi\ FWHM at  the emission peak of each cloud derived from the Gaussian fit.
The median FWHM is 24 \kms (dashed vertical line).
{\it Lower Right Panel:} 
The number of \hi\ clouds in the current sample as a function of \vlsr.
For this paper we consider only FB clouds with $|\vlsr| \gtrsim 130$ \kms. 
The number of clouds increases with decreasing $|\vlsr|$. 
}
\label{fig:properties}
\end{figure*}

The upper left panel of Figure \ref{fig:properties} shows a histogram of the measured \TL\ averaged over a $10\arcmin \times 10\arcmin$ region at the brightest part of each cloud, while the  
upper right panel  shows  \nhi.
The distributions of \nhi\ and \TL\ are quite similar, reflecting the fact that the FWHM of the clouds does not seem to be correlated with their \vlsr, their location within the FB, or their distance from the Sun (see Section~\ref{sec:detectability}). 
The fact that clouds are detected in increasing numbers down to the sensitivity limit of the survey implies that many more \hi\ clouds would be detected through more sensitive observations.
Statistical properties of the \hi\ cloud population are summarized in Table \ref{tab:cloud_statistics}.

\begin{deluxetable*}{cccc}
\label{tab:cloud_statistics}
\tablecolumns{4}
\tablewidth{0pt}
\tablecaption{Properties of the 228 FB \hi\ Clouds \label{:tab:cloud_statistics}}
\tablehead{
\colhead{Property}&
\colhead{95\% range}&
\colhead{Mean}&
\colhead{Median} \\
      }
\startdata
$\nhi  \ (10^{19} \ \cmm)$ & 0.22 -- 7.81 & 1.82 & 0.95 \\
$\nhi~\text{error} \tablenotemark{a} \ (10^{19} \ \cmm)$ & 0.025 -- 0.16 & 0.075 & 0.05 \\
FWHM (\kms) & 11.7 -- 46.6 &  25.8 & 24.2 \\
\TL  \ (K) & 0.06 -- 1.61 &  0.37 & 0.20 \\
$\TL / (\TL~\text{error} )$\tablenotemark{a} & 6.0 -- 77.7  &  25.2  & 17.9 \\
\enddata
\tablenotetext{a}{Errors are  $1\sigma$ from the Gaussian fit.}
%\vspace{-0.5cm}
\end{deluxetable*}

The lower right panel of Figure \ref{fig:properties} shows the distribution of the cloud velocities, a key quantity for our purposes.  
The number of clouds at both positive and negative \vlsr\ increases as $|\vlsr|$ decreases, down to our velocity cutoff of $\sim\pm130$ \kms.
There are also more clouds at negative velocity (124) than at positive velocity (104) by a ratio of $54\%$ to $46\%$.

\begin{figure}
\includegraphics[width=0.47\textwidth]{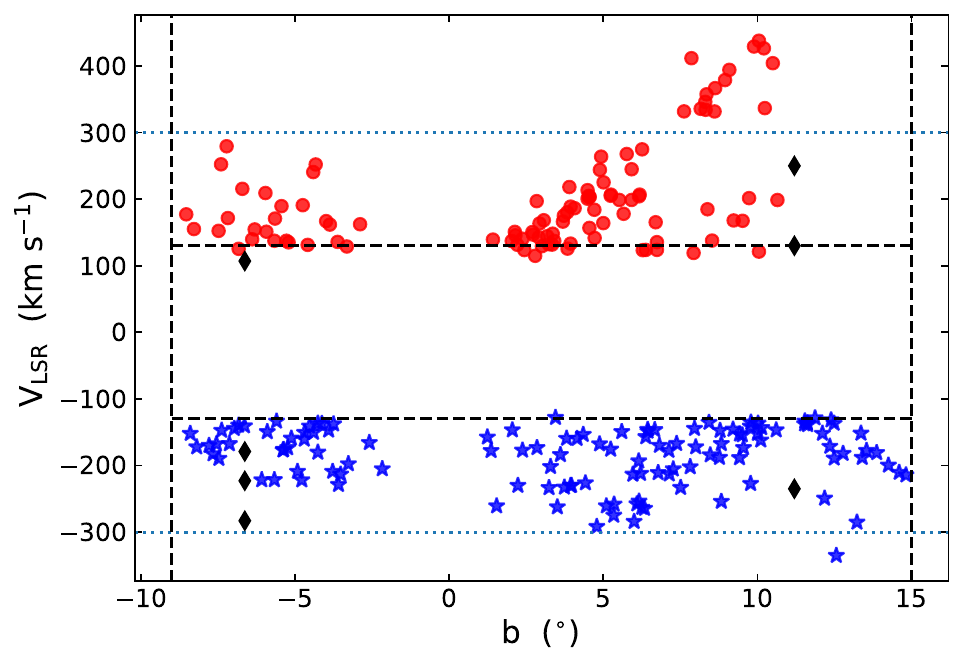}
\caption{
The \vlsr\ of the GBT clouds  shown as a function of latitude. 
Vertical dashed lines show the latitude limits of the GBT survey.
The absence of clouds within a few degrees of the Galactic plane reflects the survey limits.
The absence of clouds at $|\vlsr| \lesssim 130$ \kms\ is our choice to concentrate on clouds with the greatest information about population kinematics.  
The lack of positive-velocity clouds at $b \geq 10.6^{\circ}$ is a real feature of the data and cannot arise from any observational selection effect. 
The region lacking clouds corresponds to a distance from the Galactic plane of $z >2$ kpc.   
Black diamonds show the location and velocity of ionized components detected in UV absorption lines \citep{Fox+15,Savage+17}.
}
\label{fig:clouds_VLSR_b_UV}
\end{figure}

Cloud velocities as a function of Galactic latitude are shown in Figure \ref{fig:clouds_VLSR_b_UV}.
To our knowledge, the clouds in this sample have the highest velocity of any \hi\ emission detected thus far in the Milky Way disk, spanning $-335 \, \kms\  \leq \vlsr \leq +438 \, \kms$.
Note also that the UV absorption line components (black diamonds) have kinematics similar to the \hi\ gas in their direction.

There are two outstanding features of this figure:
\begin{itemize}

\item  Negative \vlsr\ clouds (blue stars) are found at all latitudes covered by the survey, $-9\dg \leq b \leq 15^{\circ}$, while positive velocity clouds (red circles) are restricted to $b \leq 10.6^{\circ}$. 

\item Positive velocity clouds are found to a much higher $|\vlsr|$ than negative velocity clouds by ${\sim100~\kms}$.

\end{itemize}

As we will see in Section \ref{sec:comparison_with_obs}, both of these features arise naturally from an accelerating outflow limited to distances from the Galactic plane $z \leq 2$ kpc.

Although there is highly ionized gas detected in UV absorption spectra to a slightly higher latitude, its velocities  suggest that this material also lies at $z < 2$ kpc.

\subsection{North-South Asymmetries}
\label{sec:North-South}
The FBs consist of two, nearly-symmetric lobes with similar physical properties, but there is no reason to assume that their histories are identical or that the neutral clouds they entrain have similar characteristics.
\citet{DiTeodoro18} saw no significant difference in their sample between the northern and southern \hi\ clouds, corresponding to clouds at $b > 0^{\circ}$ and $b < 0^{\circ}$, respectively.

The analysis is complicated because of the significant difference in average sensitivity (i.e., spectral noise) between the northern and southern data and the difference in area covered (see Figure~\ref{fig:noisemap}).  
To allow comparison between the two regions, we examined a subsample of 58 northern clouds at $350^{\circ} \leq \ell \leq +16^{\circ}$ and $3.5^{\circ} \leq b \leq 9^{\circ}$,  with 
$\TL > 0.11$ K.
This is similar to the sky coverage of the 68 clouds in the southern sample and the brightness of the faintest \hi\ line there (0.12 K).
We find no significant difference in the kinematics between the two sets of \hi\ clouds, confirming the conclusion of \citet{DiTeodoro18} with the newer data.
There are also no significant differences in the statistical properties of \TL\ or \nhi\ when differences in survey sensitivity over the two areas are taken into account.

%%%%%%%%%%%%%%%%%%%%%%%%%%%%%%%%%%%%%%%%%%%%%%%%%%%%%%%%
%%%%%%%%%%%%%%%%%%%%%%%%%%%%%%%%%%%%%%%%%%%%%%%%%%%%%%%% Geometry
\section{Geometry of the Fermi Bubble Observations}
\label{sec:geometry}

\begin{figure}
\includegraphics[width=0.47\textwidth]{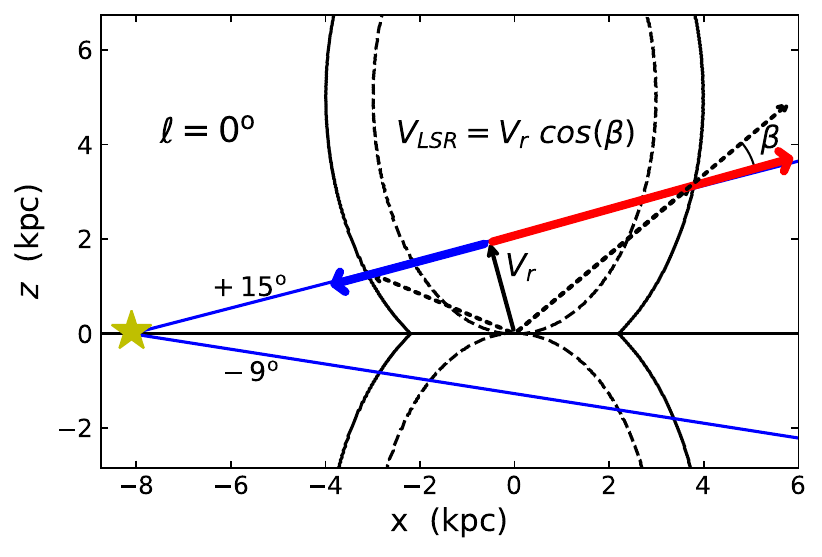}
\caption{
The FB geometry from the MB16 model of X-ray emission projected on the $x$-$z$ plane, 
where $x$ is along the line from the Sun to the GC and $z$ is distance from the Galactic plane. 
The GBT survey presented here covers latitudes between $-9^{\circ} \leq b \leq +15^{\circ}$, indicated by the solid lines  from the location of the Sun (yellow star).
Radially outflowing gas originating at the GC and observed at $b = 15\dg$ is illustrated by three arrows that mark fiducial points in the resulting LSR  velocities.
The solid black arrow marks the location where the outflow is perpendicular to the sightline and hence projects to a \vlsr\ of  zero. 
This is the outflow ``tangent point" and divides the sightline into positive and negative \vlsr\ regions.
}
\label{fig:geometry_x-z}
\end{figure}

The basic geometry of the FB as seen from the Sun is illustrated in Figure \ref{fig:geometry_x-z}.
As in Figure~\ref{fig:clouds_l-b-VLSR}, the solid and dashed black lines delineate the outer shock and inner-filled bubble from the MB16 model, respectively.
The original MB16 model includes a slight tilt to the bubbles, which we ignore.
We use a coordinate system centered on the GC where 
Galactic longitude $\ell = 0^{\circ}$ lies along the $x$-axis; the $y$-axis is perpendicular to the Sun-center line, and $z$ is distance from the Galactic plane.
The Sun, indicated as a yellow star, is at $x = -R_{0}$ where $R_{0} = 8.275$ kpc \citep{GRAVITY2021}. 

It can also be useful to employ a right-handed cylindrical coordinate system where $R$ is the distance from the GC at $z = 0$, and the azimuthal angle $\theta$ is measured from the Sun-center line at $x > 0$ increasing in the same sense as longitude, so that as $R \rightarrow \infty, \ \theta \rightarrow \ell$.
In this system, Galactic rotation has a negative $V_{\theta}$.
Distance from the GC is given by the spherical radius $r = (R^2 + z^2)^{1/2} = (x^2 + y^2 + z^2)^{1/2}$, while the distance from the Sun is $\dsun = ((x - R_0)^{2} + y^{2} + z^{2})^{1/2}$.
For simplicity, we assume a purely radial wind, such that the outflowing gas moves outward from the nucleus with radial velocity \Vr. 

Within the FBs, the amplitude of the observable \vlsr\ is determined by cosine of the angle $\beta$ between the outflow velocity, \Vr, and the line of sight: 
$\vlsr = \Vr \cos(\beta)$. 
The maximum projection occurs at the edges of the bubbles.

\subsection{The relationship between \vlsr\ and the outflow velocity \Vr}
\label{sec:TL_vs_Vrho}

\begin{figure*}
\centering
\includegraphics[width=0.48\textwidth]{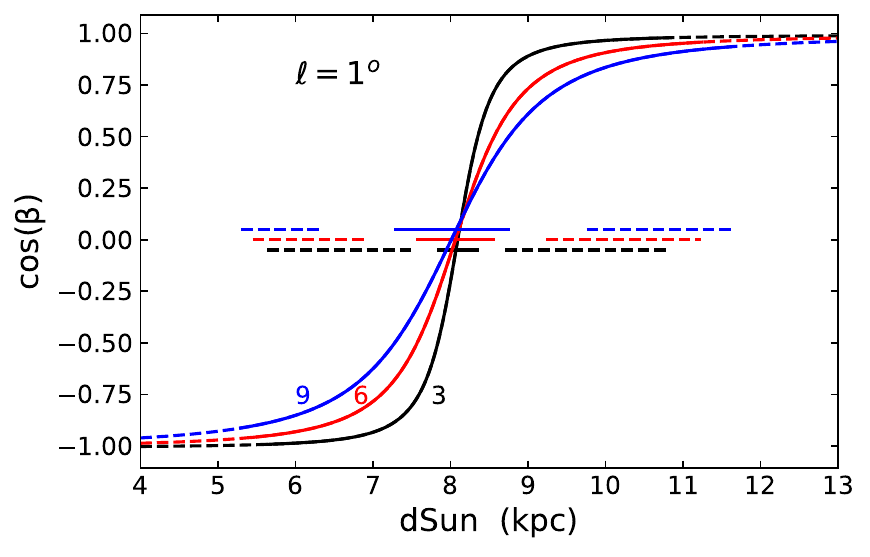}
\includegraphics[width=0.46\textwidth]{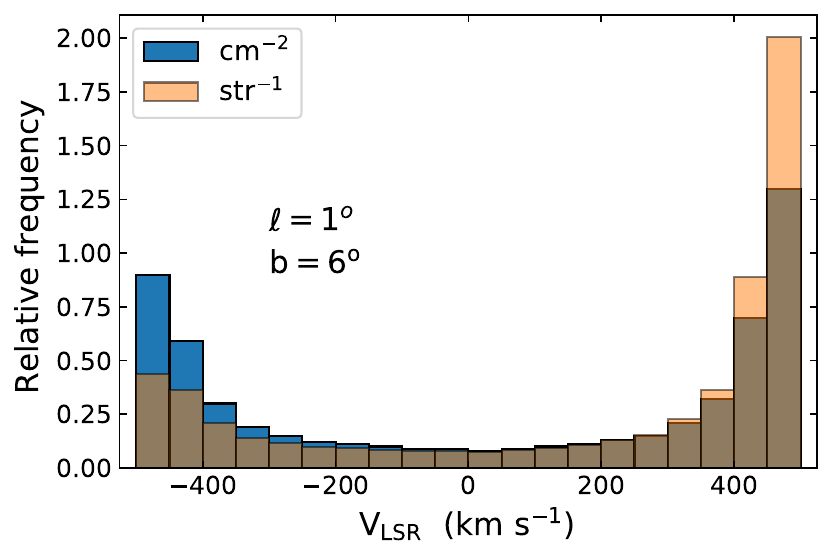}
\caption{ 
{\it Left panel:} 
The  projection factor cos($\beta$) 
at $\ell = 1^{\circ}$ for three different latitudes, ${b = 3^{\circ}}$ (black curve), $b = 6^{\circ}$ (red curve), and $b = 9^{\circ}$ (blue curve) plotted vs.\ distance from the Sun.
The observed \vlsr\ of a radial outflow is $\Vr\cos(\beta)$. 
Curves are solid lines for locations within the FBs and broken outside.
Horizontal solid lines show the range of distance over which $|\cos(\beta)| < 0.5$ and  dashed lines show the range for $|\cos(\beta)| > 0.8$.
Note that a larger range of distance is covered by the largest values of the projection factor whereas velocities near zero are found only in a limited region.
{\it Right panel:}
The relative column density (blue) and volume density (orange) for a uniform velocity outflow with $\Vr = 500$ \kms\ when observed at ($\ell, b) = (1\dg,6\dg)$.
Because of the velocity crowding shown in the left panel, extreme velocities with $|\vlsr| \approx \Vr$ are favored. 
For a uniform density of gas along this line of sight, spectra should resemble one of these distributions.
}
\label{fig:model_cosbeta_vs_dsun}
\end{figure*}

While the outflow velocity cannot be observed directly, there are cases that can provide interesting limits on this quantity.
The solid black arrow in Figure~\ref{fig:geometry_x-z} marks the location where the outflow is perpendicular to the sightline and hence projects to a \vlsr\ of  zero. 
 This is the outflow ``tangent point".
Towards the Sun (to the left in the Figure) the outflow projects as a negative \vlsr; to the right, away from the Sun, as a positive \vlsr.

Along any sightline, negative and positive velocities arise in different regions of the FB, whose distance from the Galactic plane  may vary by a factor $\geq 2$.
The volume sampled by negative velocities is always smaller than positive velocities except exactly at $b = 0\dg$.
This figure illustrates important selection effects in the current latitude-limited survey: the data cover distances from the plane of $z \lesssim 3$ kpc in redshifted emission but only $z \lesssim 1$ kpc in blueshifted emission.

The effects of the varying projection of \Vr\ onto \vlsr\  can be understood from Figure~\ref{fig:model_cosbeta_vs_dsun} (left panel). Here $\cos(\beta)$ is plotted as a function of \dsun\ for several latitudes  at $\ell = 1^{\circ}$.
One consequence of the geometry of the projected outflow  is that higher values of $|\cos(\beta)|$ are favored over values closer to zero because of the limited length of the path through the FB where $\beta \approx 90^{\circ}$.
This is illustrated  by the horizontal lines in the center of the left panel.
Three solid horizontal lines cover the range of distance through the FBs  where $|\cos(\beta)| < 0.5$
and three dashed horizontal lines show locations where  
$|\cos(\beta)| > 0.8$, a projection factor which gives $|\vlsr| \approx \Vr$.

At low longitudes there is a much larger path through the FBs with $|\vlsr| \geq 0.8 \Vr$ than with  $|\vlsr| \leq 0.5 \Vr$ by factors ranging from 8 at $b = 3\dg$ to 2 at $b = 9\dg$.
This ``velocity crowding" -- a circumstance in which there is a nearly constant \vlsr\ over a relatively large distance  --  is a common occurrence in the Galaxy \citep[e.g.,][]{Burton1971}. 
Its effect on the observable \vlsr\ is shown in the right panel of Figure \ref{fig:model_cosbeta_vs_dsun} towards the sightline $\ell = 1^{\circ}$, $b = 6\dg$ for an outflow with a constant $\Vr=500~\kms$.

If  the gas density is uniform along this path and the outflow velocity is constant,  the resulting emission or absorption spectrum will be similar to the histogram shown in blue. 
Gas piles up at the extreme projected velocities where cos($\beta$)  changes slowly with distance.  
If we display instead the volume subtended in a given solid angle along this path, the result is shown in orange.  
The fact that the \hi\ clouds are most numerous at low $|\vlsr|$  and not high $|\vlsr|$ (Figure~\ref{fig:properties}) thus carries significant information on the density and velocity structure of the neutral outflow:  \Vr\ must vary throughout the FBs.

%%%%%%%%%%%%%%%%%%%%%%%%%%%%%%%%%%%%%%%%%%%%%%%%%%%%%%%%%%
%%%%%%%%%%%%%%%

\section{Limits on the outflow velocity}
\label{sec:Vout_limits}

\begin{figure}
\includegraphics[width=0.46\textwidth, trim= 0 0 1.5cm 1cm, clip]{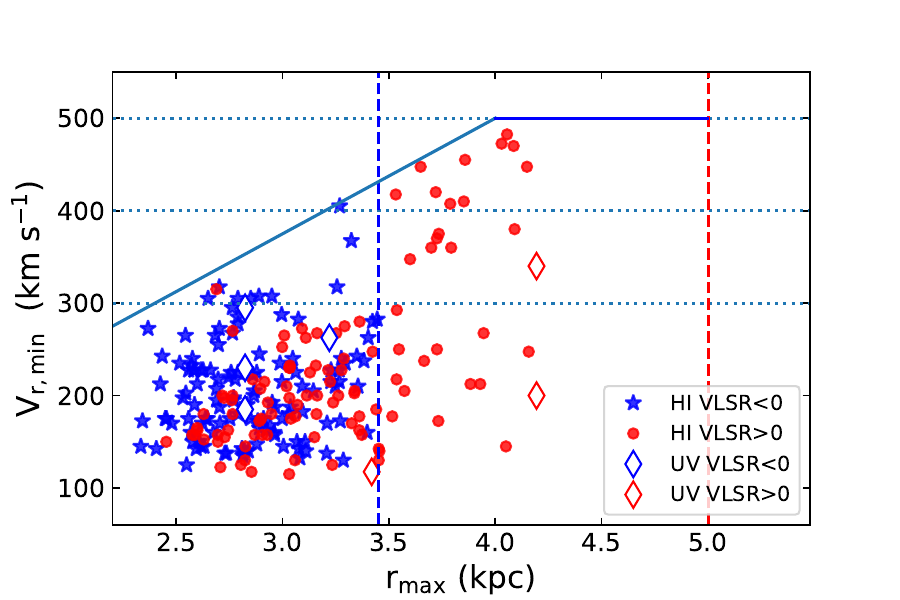}
\caption{
The minimum outflow velocity \Vr\ required to produce the observed \vlsr\ of \hi\ clouds in the FBs.
This is derived by placing every cloud in the GBT survey at the nearest FB wall in its direction, the location where $|\cos(\beta)|$ is greatest.
The dashed blue and red lines show the limits of the GBT survey for the near and far FB walls, respectively.
The solid blue line  shows the velocity adopted for  outflow model M-1: 
$\Vr = 125 \ r \  \kms \ \mathrm{kpc^{-1}}$ for $r \leq 4$ kpc, and  $\Vr = 500\ \kms$ at $r \geq 4$ kpc.  
}
\label{fig:clouds_Voutmin_rho_UV}
\end{figure}

From the discussion in Section \ref{sec:geometry} and especially Figure~\ref{fig:model_cosbeta_vs_dsun}, it is clear that the greatest projection of an outflow velocity  onto \vlsr\ occurs at  the FB walls.  
The walls are also at the greatest distances from the GC along any line of sight through the FBs.
This means that the measured \vlsr\ of every cloud can be used to determine a lower limit on \Vr\ under the assumption that the cloud is located at the FB wall in its direction -- the near wall for $\vlsr < 0$ and the far wall for $\vlsr > 0$.
Figure \ref{fig:clouds_Voutmin_rho_UV} results from placing each cloud in the sample at the nearest FB wall in its direction, and thus at a known maximum distance $r_\mathrm{max}$ from the GC, then calculating the minimum radial velocity \vrmin\ necessary to produce the observed \vlsr.

There are two important features of Figure~\ref{fig:clouds_Voutmin_rho_UV}:
first, the minimum \vrmin\ increases with $r_\mathrm{max}$, an outcome that cannot be attributed to simple projection effects, as the projection of \Vr\ on \vlsr\ is greatest at low latitude and low longitude and thus low $r_\mathrm{max}$.
This implies that \Vr\ must increase with $r$, as already suggested by \citet{Lockman2020}. 
An alternative possibility is that close to the GC the clouds lie in a narrow cone within the FBs, where the projection is unfavorable and thus $\Vr \gg \vrmin$.
However, as the clouds span a wide range in longitude consistent with their location in a filled FB cone (Figure~\ref{fig:clouds_l-b-VLSR}) this possibility would require an elliptical outflow with the Sun located preferentially along the short axis.

The second important feature lies in the difference between clouds assigned to the near (blue stars) and far (red circles) FB walls.  
Whereas clouds in the near side of the FBs are seen to the vertical limit of the survey (dashed blue line), there is an empty region at $r_\mathrm{max} \geq 4.2 $ kpc where clouds could have been detected on the far side of the FBs, but are not.
This void is also apparent in the raw data of Figure~\ref{fig:clouds_l-b-VLSR} and Figure~\ref{fig:clouds_VLSR_b_UV}. 
As noted in Section~\ref{sec:cloud_properties}, there are no \hi\ clouds with $\vlsr \gtrsim 130$ \kms\  at $b > 10.6^{\circ}$.

From Figure~\ref{fig:clouds_Voutmin_rho_UV} we derive the simple model for \Vr\  shown by the slanted line: $\Vr(r) = 125 \  r$  $\kms \, \text{kpc}^{-1}$. 
This model is not constrained by the current data at $r \lesssim 1.5$ kpc, nor at $r \gtrsim 4$ kpc.
We assume that \Vr\ is constant at $500\ \kms$ for $r > 4$ kpc, though in practice this has no effect on our analysis. 
We will refer to this $\Vr(r)$ as outflow model M-1.
An accelerating outflow of this kind has  already been proposed by \citet{Lockman2020} based on similar evidence, but with a slightly different acceleration law and maximum velocity. 

 Note that because the projection factor $\cos(\beta)$ varies slowly at the location of the FB walls (Figure~\ref{fig:model_cosbeta_vs_dsun})  the values of $\vrmin$ are robust to deviations in  the exact location of each cloud.  
When the data are analyzed as if each cloud were located 0.5 kpc inside the FB wall,   the major effect is to shrink the outflow slightly in $r$.
A gradient in \Vr\ is still required, with a similar but slightly larger magnitude than in outflow model M-1.

%%%%%%%%%%%%%%%%%%%%%%%%%%%%%%%%%%%%%%%%%%%%%%%%%%%
%%%%%%%%%%%%%%%%%%%%%%%%%%%%%%%%%%%%%%%%%%%
%%%%%%%%%%%%%%%%%%%  Variation of Cloud properties 
\section{Variation of cloud properties \\within the Fermi bubbles}
\label{sec:variation}

\subsection{The variation of \TL\ and \nhi\ with $r$}
\label{sec:TL_rho}

Using the model for $\Vr$ from the previous section, an estimate of the 3D spatial location of each cloud in the GBT sample can be derived from its observed $\ell$, $b$, and \vlsr. 
One interesting result arises immediately: a systematic variation in the observed \TL\ with the derived distance from the GC, $r$.

\begin{figure*}
\includegraphics[width=0.98\textwidth]{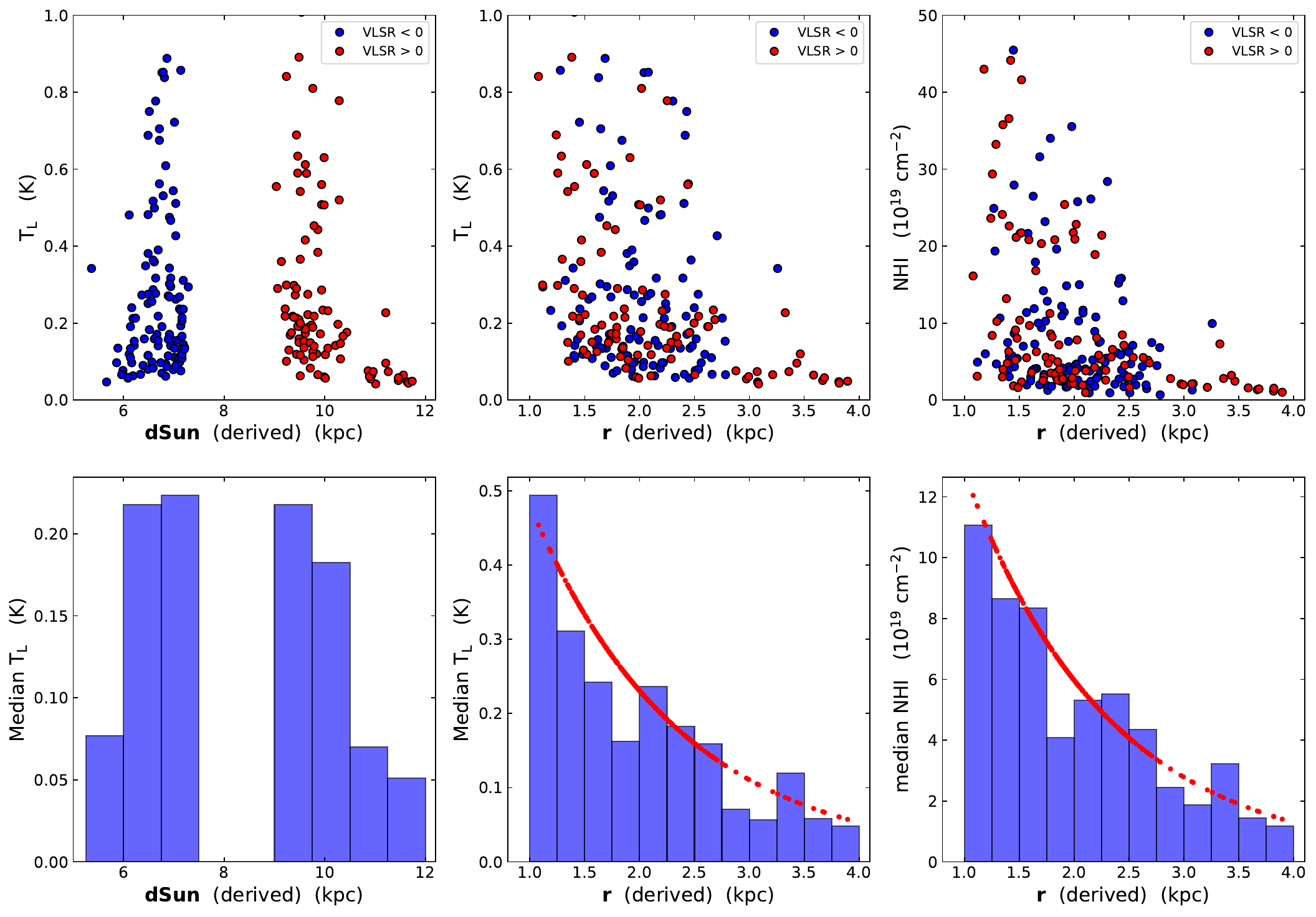}
\caption{
Properties of the GBT clouds as derived from outflow model M-1.
The upper panels show \TL\ and \nhi\ values for individual clouds and the lower panels show the binned median distributions.
There is no significant dependence on $\dsun$ (left panels) but there is a dependence on distance from the Galactic center, $r$, shown in the lower center and right panels.
The curves are exponential fits to the individual values and are evaluated at the $r$ of each cloud.
The fits are essentially identical for $\nhi(r)$ and $\TL(r)$. This is likely evidence of cloud evolution with distance from the Galactic center.
}
\label{fig:TL_hists}
\end{figure*}

The dependence of \TL\ on both \dsun\ and $r$ is shown in Figure \ref{fig:TL_hists} for the individual clouds (top panels) and binned median values (bottom panels), for distances derived using outflow model M-1.
There is no strong change in \TL\ with distance from the Sun  out to $\dsun \approx 10$ kpc.
The effect with distance from the GC, however, is  a monotonic decrease in the median \TL\ with $r$,  
seen even when the populations are divided into positive and negative \vlsr\ samples (not shown here).
Values of \TL(r)\ are  fit by an exponential curve of the form ${\TL = 0.8 \exp(-0.74r)}$ K, with $r$ in kpc, as shown by the red curve.
The density of points along the curve shows the number of clouds at each derived $r$.

An essentially identical result is obtained for the variation in \nhi\ with $r$, which is to be expected, given that $\nhi \propto \TL\ \times {\rm FWHM}$, and there is no systematic variation of FWHM with $r$. 
Cloud values of \nhi\ are fit with an exponential 
${\nhi = 20.2 \times 10^{19} \ \exp(-0.76r)}$ \cmm.

%%%%%%%%%%%%%%%%%%%%%%%%%%%%%%%%%%%%%%%%%%%%% TL_vs_dist_rho_4 

%%%%%%%%%%%%%%%%%%%%%%%%%%%%%%%%%  TL vs rho
\begin{figure*}
\centering
\includegraphics[width=0.45\textwidth]{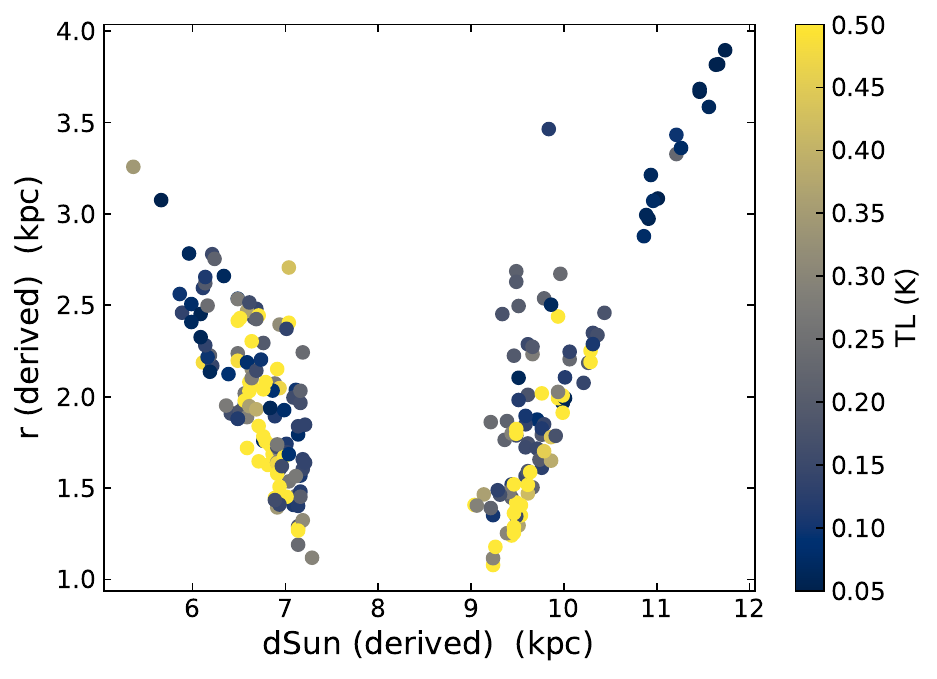}
\hspace{10pt}
\includegraphics[width=0.45\textwidth]{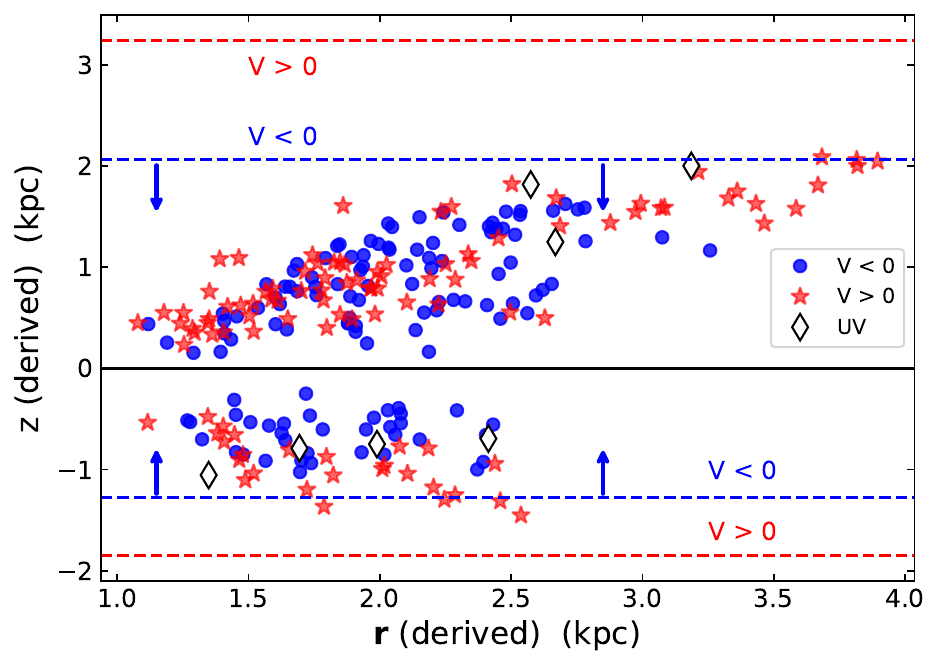}
\caption{ {\it Left panel:}
The distance from the GC, $r$, vs.\ distance from the Sun, $\dsun$, for the survey clouds derived using outflow model M-1.
The color is proportional to the observed \TL, with yellow  for the brighter lines and black for the fainter.
It is clear that the 21cm line brightness is most strongly related to $r$ rather than $\dsun$.
{\it Right panel:}
The distance from the Galactic plane, $z$, vs. distance from the GC, $r$, for the survey clouds  analyzed using outflow model M-1.
Dashed red lines at $z = -1.8 \ and \ +3.2$ kpc show the maximum distance from the Galactic plane of the far walls of the FBs covered by the Southern and Northern data.
From the geometry alone, clouds at $\vlsr > 0$ could have been detected up to these values of $z$.
The blue dashed lines show values of $z$ for the tangent points at the latitude limits of the survey (see Figure~\ref{fig:geometry_x-z}).
Clouds with $\vlsr < 0$ must be located closer to the Sun than the tangent points and thus the blue dashed lines are shown as limits.
Clouds at  $z \lesssim 2$ kpc span a range of $r$ and it  appears that the distance from the Galactic plane limits the extent of the cloud population,  rather than distance from the GC.
The ionized lines detected in UV absorption (open diamonds) have kinematics similar to that of the \hi.
}
\label{fig:TL_rho}
\end{figure*}

%%%%%%%%%%%%%%%%%%%%%%%%%%%%%%%%%%%%%%%  sigma TL(rho)
\begin{figure}
\includegraphics[width=0.46\textwidth]{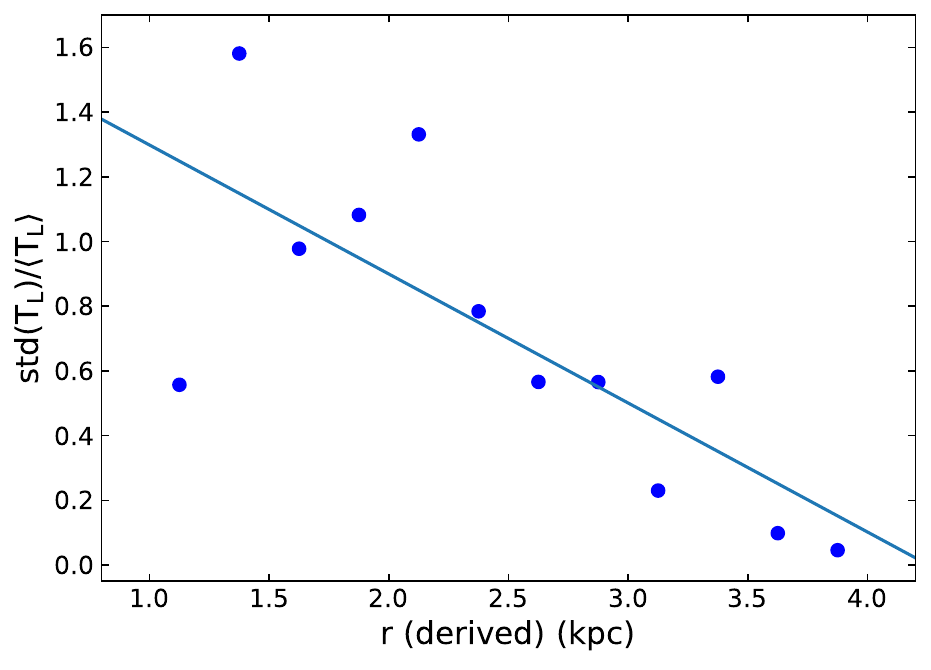}
\caption{
The ratio of the standard deviation about the mean \TL\ to the mean \TL\ calculated over intervals of  distance from the GC  derived from outflow model M-1. 
There is a significant decline in 
${\rm std(\TL) / \langle\TL\rangle}$ with  distance from the GC, implying that the cloud population is becoming structurally more uniform with distance from the GC.
}
\label{fig:sigma_over_mean}
\end{figure}

While there is almost certainly some systematic change of \TL\ with \dsun (see Section~\ref{sec:TL_vs_dSun}), any such effect in the complete GBT sample seems to be small compared to the change  with distance from the GC, and is probably washed out by changes in the intrinsic cloud structure within the FBs. 
This is very evident from the left panel of Figure~\ref{fig:TL_rho}, where line brightness is shown colored from black to yellow on a plot of $r$ vs.\ \dsun.
Bright clouds (yellow) are found over the range $6 \leq \dsun \leq 10$ kpc, whereas faint clouds (black) dominate at $r > 2.5$ kpc, regardless of their distance.

The right panel of Figure~\ref{fig:TL_rho} shows the derived distance from the Galactic plane as a function of distance from the GC for the \hi\ clouds and the UV absorption components as derived from outflow model M-1.
It is apparent that clouds at a given distance from the plane, $z$, can be found over a wide range of $r$.
The absence of clouds at $z \gtrsim 2$ kpc thus reflects a limit set by distance from the Galactic plane, not distance from the Galactic center.

%%%%%%%%%%%%%%%%%%%%%%%%%%%%%%%%%%%%%%%  sigma TL(rho)
\subsection{Decrease in the scatter of \TL\ with $r$}
\label{sec:decrease_scatter}

The broad trends in $\TL(r)$ and $\nhi(r)$ shown in the lower panels of Figure~\ref{fig:TL_hists} lie amidst an enormous variation in \TL\ from cloud to cloud at most values of $r$ and \dsun. 
The magnitude of the variation can be characterized by the ratio  ${\rm std(\TL)/\langle\TL\rangle}$: the standard deviation about the mean \TL\ as a fraction of the mean \TL.
This is shown in Figure \ref{fig:sigma_over_mean}  for binned data  plotted against $r$.  
The correlation with $r$ is significant with a slope of $-0.40\pm0.11 \ (1\sigma)$ (Spearman rank coefficient $\rho_\mathrm{S}=-0.671$, $p=0.017$).
In contrast, there is no significant correlation with distance from the Sun ($\rho_\mathrm{S}=-0.38$, $p=0.28$).

The relative dispersion falls by an order of magnitude from 
$\approx 1$ close to the GC to $\approx 0.1$ for clouds most distant from the center. 
The most reasonable explanation for the decrease in both the median \TL\  and its relative dispersion  with  $r$ is an evolution of clouds within the FB flow -- clouds become more diffuse with less internal structure in their \hi\ as they move outward in the nuclear wind.
This statistical result is consistent with the change in morphology of FB \hi\ clouds as seen in high-resolution maps and studies of their molecular content \citep{DiTeodoro2020,Noon2023,DiTeodoro2026a}. 

%%%%%%%%
%%%%%%%%%%%%%%%%%%%%%%%%%%%%%%%%%%%%%%%%
%%%%%%%%
\section{Observational effects on Cloud Detectability}
\label{sec:detectability}

In this section we examine factors that affect our ability to detect \hi\ clouds in the GBT survey, factors that must be included in any simulation of the observations.

The detectability of a cloud is proportional to $\TL / \sigma$, the ratio of the line brightness temperature to the noise in a spectrum, $\sigma$.
In the previous section we found an effect that is intrinsic to the \hi\ clouds, namely a decrease in \TL\ with distance from the GC.  
Here we summarize two effects rooted in the data: first, any  distance-dependence of \TL\ owing to the fixed angular resolution of the GBT, 
and second, the varying noise level across the observed  region, which makes the survey more sensitive to some parts of the FBs than to others. 
Details of this analysis are given in Appendix~\ref{app:TL-vs-dSun}.

\subsection{Variation of $T_{L}$ with distance from the Sun}
\label{sec:TL_vs_dSun}

The \hi\ clouds in the FBs are generally not point sources to the GBT beam, but are extended with considerable internal structure \citep{DiTeodoro18,Noon2023}.
We study their distance dependence using a subsample of clouds derived from the most sensitive region of the survey, the area bound by $|\ell| \leq 5\dg$ and $5\dg \leq \ b \leq 15\dg$, which we call the ``deep" region.
From the geometry of the FB it is clear that clouds in the deep region with $\vlsr < 0$, i.e., those on the near side of the FBs, must come from a relatively small range of \dsun.
Using the outflow model M-1 from Section \ref{sec:Vout_limits}, the 39  clouds in the ``deep-near" region have a derived mean $\dsun = 6.3\pm0.4$ kpc. 
The binned median \TL\ for these clouds can be described approximately by an exponential distribution with a median of 0.15 K and a scale of 0.13 K (see Appendix \ref{app:TL-vs-dSun}).

We estimated the distance-dependence of \TL\ for these clouds by averaging the \hi\ data over regions of different angular size to simulate placing the cloud at a larger distance.
The results (Appendix \ref{app:TL-vs-dSun}) yield an empirical relationship $\TL \propto 6.3 / \dsun$ with \dsun\ in kpc, where the factor 6.3 is the mean distance to the deep-near sample derived from outflow model M-1. 

While this distance-dependence is fairly mild, these results suggest that about half of the 39 clouds in the deep-near sample would not be detectable at the greatest  distances within the FB volume covered by the GBT survey.
This will be taken into account in simulations of the data in Section \ref{sec:models}.

%%%%%%%%%%%%%%%%%%%%%%%%%%%%%%%%%%%%%%%%%%%%%%%%%%
%%%%%%%%%%%%%%%%%%%%%%%%%%%%%%%%%%%%%%%%%%%%%%%%%%

\subsection{Variable noise over the GBT survey region}

The exact noise of the GBT survey data is given in Figure~{\ref{fig:noisemap}.
For the simulations, the survey area was broken into 15 regions within which the noise in a 3 \kms\ channel was taken to be constant, ranging from 20 mK for the deep field, to $\geq 50$ mK in a few small areas.
The simulations require that the model clouds have a \TL\ that is some factor (usually three) times the survey noise in its direction.
The variability of noise means that the location of clouds ``observed" in the simulation will not necessarily reflect their actual spatial distribution, but they can be faithfully compared to the GBT data, which is subject to the same systematic effects.

%%%%%%%%%%%%%%%%%%%%%%%%%%%%%%%%%%%%%%%%%%%%%%%%%%%%%
%%%%%%%%%%%%%%%%%%%%%%%%%%%%%%%%%%%%%%%%%%%%%%%%%%%%%
%%%%%%%%%%%%%%%%%%%%%%%%%%%%%%%%%%%%%%%%%%%%%%%%%%%%%
\section{Kinematic Models}
\label{sec:models}

Given the complex relationship between the observables --  
$\ell$, $b$, and \vlsr\ -- and conditions at specific locations within the FBs, it is useful to simulate the data with simple kinematic models (for more complex, physically-motivated wind models, see \citealt{Afruni+2026}). 
Our primary goal is to understand the basic kinematic structure of the nuclear outflow as it is observed in the 21cm line.  

Each velocity channel of every spectrum in our survey is a measurement of a unique location within the FBs.
The data show that spectra  at nearby positions can have  emission  components that span a range of velocities, which is not consistent with FB models where the material is confined to a thin shell.
The UV absorption spectra through the FBs \citep{Fox+15,Bordoloi+17,Savage+17}, in fact, detect multiple velocity components  over an infinitesimal solid angle.
The data thus suggest that we use FB models with gas  flowing outward into filled cones.

In many past analyses of gas kinematics in the FBs, it was  assumed that the outflow was collimated with  a constant opening angle.  
Derived values ranged from $\leq110\dg$ to $160\dg$ \citep{Keeney2006,McClure-Griffiths+13,Bordoloi+17,DiTeodoro18,Lockman2020}.
In this work, we have chosen to model the outflow as isotropic, but confined within the outer walls of the FBs,  which extend over a considerable area at every $z$, even near the Galactic plane.
This is motivated by the observation that the FBs  have altered the structure of the interstellar medium for several kpc around the GC, even close to the Galactic plane \citep{Lockman84,Lockman_McClureGriffiths16}.

\subsection{Parameters of the simulations}
\label{subsec:parameters}

The models consist of many test particles (``clouds")  within the MB16 boundaries of the FBs.
Following the results of Section~\ref{sec:Vout_limits} (see Figure~\ref{fig:clouds_Voutmin_rho_UV}), we adopt  a fiducial model outflow of the form $\Vr = 125 \ r$ \kms\ kpc$^{-1}$ for $r\leq4$ kpc and  $\Vr = 500 \, \kms$ for $r\geq4$ kpc (model M-1).
It is also useful to consider a model with a constant outflow velocity, model M-0, where $\Vr = 500$ \kms\  independently of $r$, to show effects that are purely geometrical. 
In model M-0, the observed \vlsr\ is directly proportional to $\cos(\beta)$ and it thus maps the projection factor throughout the FB volume, weighted by the survey coverage and sensitivity.

We first generate a set of clouds with a uniform space density centered on the Galactic center.
Clouds that lie outside the boundaries of the FB outer walls as given in the MB2016 model are then eliminated.
To approximate a constant mass flux outflow, clouds are then systematically eliminated until the simulation has an equal number of clouds in every interval of distance from the GC 
$\Delta r$, i.e.,  $n(r) \Delta r$ is constant within the FB boundaries. 

Clouds located within the region observed by the GBT are then selected, and outflow velocities \Vr\ are assigned using a model such as M-1.
From their location and \Vr, a value of \vlsr\ is calculated, and, to match the selection of the data, only clouds with 
$|\vlsr| \geq 130$ \kms\ are retained. 
This places a limit on the volume studied whose boundaries depend on the form of \Vr.

\begin{figure}
\includegraphics[width=0.45\textwidth]
{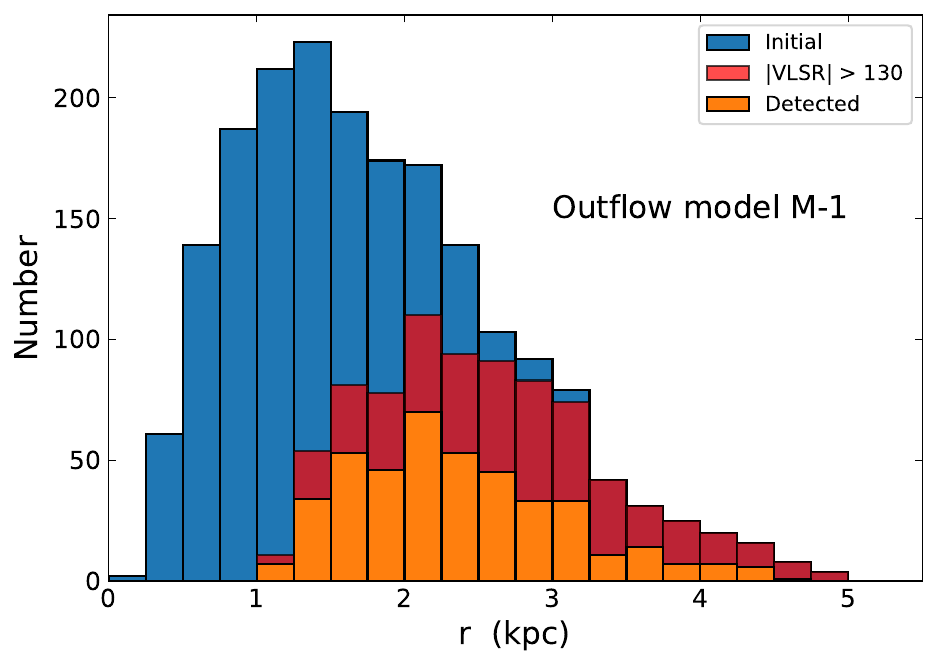}
\caption{
The  number of clouds in a typical simulation in intervals of distance from the GC, $r$.
All models initially have a constant number of clouds in intervals of $\Delta r$ within the FB boundaries which, when convolved by the GBT survey sky coverage, gives the blue histogram. 
The model is further restricted to clouds with $|\vlsr| > 130$ \kms\ evaluated with outflow model M-1, giving the  reduced distribution  shown in red.
When sensitivity and detectability criteria are included the results are shown in orange.
 These are the clouds that would be expected to be detected in the GBT survey if the outflow has a uniform spatial structure. 
The model suggests that the survey is mainly informative about  the volume of the FBs between  $1.5 \lesssim r \lesssim 3.5$ kpc.
}
\label{fig:M1_n_of_rho}
\end{figure}

Finally, each model cloud is assigned a value of \TL\ randomly drawn from an  exponential distribution with properties consistent with the ``deep-near" sample (Appendix A) whose median follows the dependence on $r$ shown in Figure~\ref{fig:TL_hists}, with a minimum value of 0.05 K.
This \TL\ is then reduced by the \dsun-dependent factor $6.3/\dsun \leq 1$ (Section~\ref{sec:TL_vs_dSun}), 
  resulting in an ``observable" \TL\ for each cloud in the simulation.   
If this value of \TL\ lies above some detectability criterion (usually three times the survey noise in the direction of the cloud) the cloud is ``detected".
The effect of the selection process  on the characteristics of simulated cloud population is illustrated in Figure~\ref{fig:M1_n_of_rho}.

This final step usually produces a significant reduction in the volume of the FBs that can be simulated, but it corresponds to real sensitivity limits in the GBT data that prevent us from sampling regions of the FBs that lie at large $r$ and \dsun.
The importance of these considerations will be  apparent in the next section.

\section{Comparison with observations}
\label{sec:comparison_with_obs}

\begin{figure}[htb!]
\includegraphics[width=0.45\textwidth]{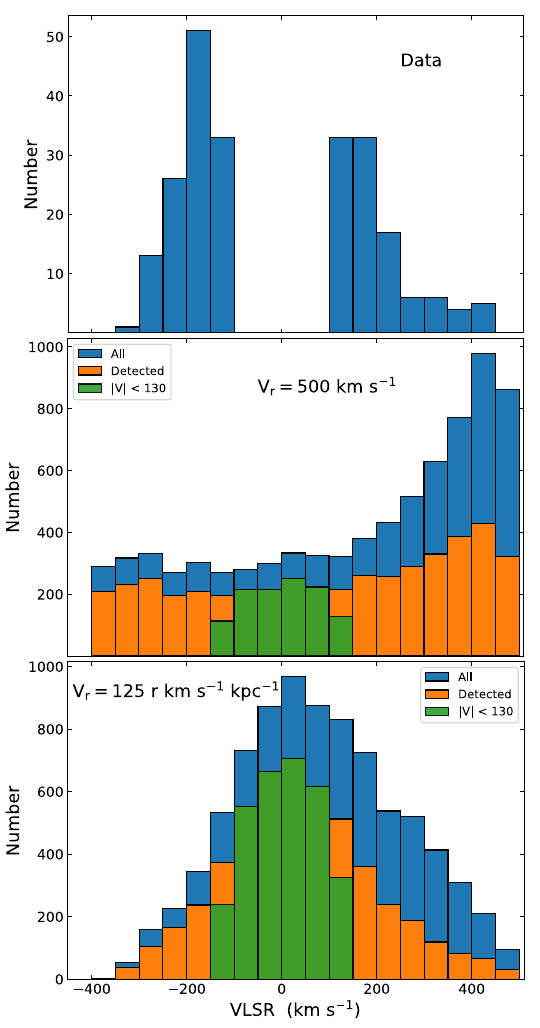}
\caption{
The observed \vlsr\ of the GBT clouds (data, top panel) and two simulations,  one for a constant $\Vr = 500$ \kms\ (center  panel)  and one for outflow model M-1 (lower panel) 
that has ${\Vr = 125 \ r \leq 500}$ \kms.
Model velocities in the range $|\vlsr| \leq 130$ \kms\ are colored in green to match the restrictions on the observational data.
For the models, the blue bars show all the simulated clouds in the GBT survey region, while the orange bars show those above the $3\sigma$ detection threshold for the current survey.  
Both the data and model M-1 have increasing numbers of clouds toward $\vlsr = 0$ while the constant velocity model M-0 has a peak in cloud numbers at high positive \vlsr\  that is not consistent with the observations.
}
\label{fig:model-n-of-VLSR-v2}
\end{figure}

\begin{figure*}[htb!]
\includegraphics[width=0.47\textwidth]{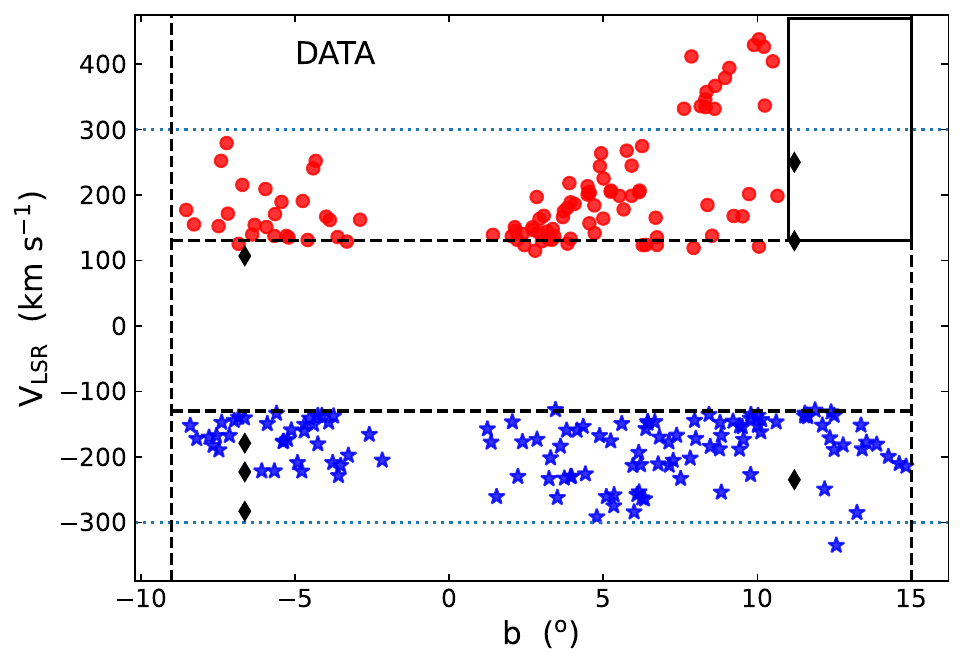}
\hspace{10pt}
\includegraphics[width=0.47\textwidth]{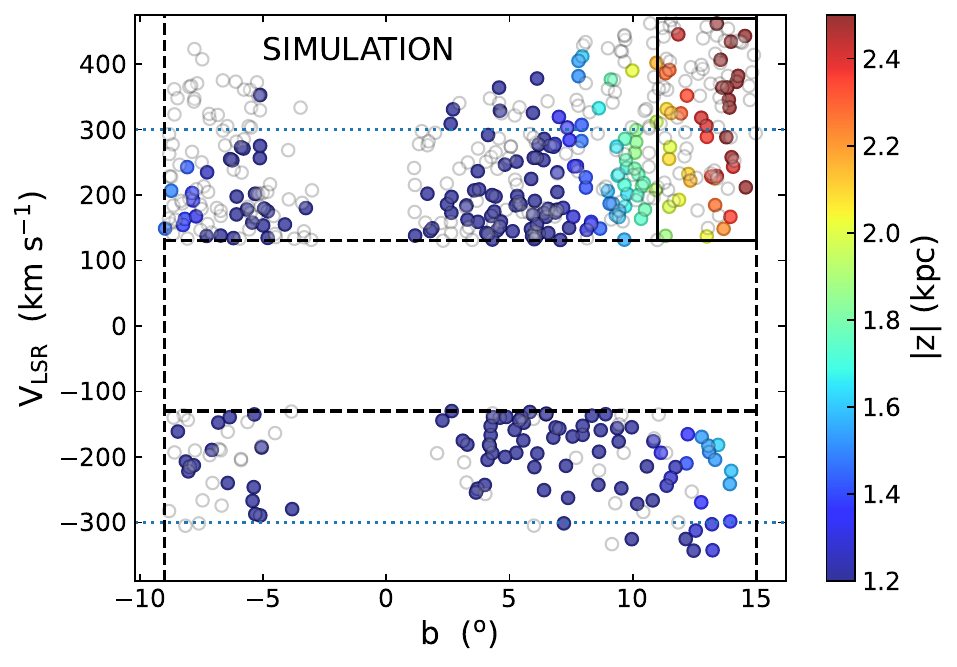}
\caption{
{\it  Left panel:}
The observed \hi\ clouds from the GBT sample plotted in \vlsr\ vs.\ latitude.
Vertical dashed lines mark the GBT survey latitude limits; horizontal dashed lines mark the velocity limit $|\vlsr| =  130$ \kms.
Black diamonds show ionized components detected in UV absorption lines, and the box at positive velocity and $b \geq 10.6\dg$ marks the region where clouds could have been detected but were not. 
This panel contains the identical data as Figure~\ref{fig:clouds_VLSR_b_UV}.
{\it Right panel:} 
Identical plot for model clouds derived from outflow model M-1.
Clouds in the full simulation are shown as open symbols; those which pass the detection limit as filled symbols colored by $|z|$.  
This illustrates the importance of observational selection effects,  especially for clouds in the most distant parts of the FBs.
Simulated clouds in the boxed area  all are at $z > 2$ kpc,  where there are no corresponding clouds in the data.
}
\label{fig:M2_VLSR_vs_b}
\end{figure*}

Figure \ref{fig:model-n-of-VLSR-v2} compares two simulations of $n(\vlsr)$ with the distribution of \vlsr\  from the observations.
The measured values of cloud \vlsr\ (top panel) are a fundamental result of the GBT survey and do not depend on a model.
The values of \vlsr\ created in the simulations (middle and bottom panels) depend directly on the adopted model for $\Vr$. 
In the simulated populations of Figure~\ref{fig:model-n-of-VLSR-v2}, the blue histograms represent the full simulated sample, while the orange histograms show the subset of clouds that would be detectable in the current GBT survey, with low-velocity clouds ($|\vlsr| \leq 130$ \kms\ that are not considered here) marked in green.

The constant outflow velocity simulation shown in the central   panel of Figure~\ref{fig:model-n-of-VLSR-v2} is largely a diagram of $\cos(\beta)$ over the volume of the survey weighted by observational selection effects.
The projection of \Vr\ onto \vlsr\ is greatest at the FB walls, and the distant walls contribute more volume to the survey area than the nearby walls, hence for a constant \Vr, $n(\vlsr)$ rises to positive velocity.
The geometry of the situation is illustrated in  the left panel of Figure~\ref{fig:model_cosbeta_vs_dsun}, and its consequences shown  in the right panel of that figure. 
When survey sensitivity is taken into account (orange), the numbers at large \vlsr\ are reduced, as many of these distant clouds then fall below the detection limit.

The bottom panel of Figure~\ref{fig:model-n-of-VLSR-v2} shows results for the same set of model clouds when \Vr\ is allowed to vary linearly with $r$ according to our outflow model M-1.
This simulation reproduces the main pattern in the data of increasing numbers as $|\vlsr|$ is reduced. This model also predicts a sharp peak in the cloud counts at $\vlsr\simeq0$ \kms, which however we cannot probe observationally due to confusion with foreground emission from the Milky Way disk.
We emphasize that cloud positions and observational selection effects are identical in the two simulations, only the values of \Vr\ differ.

The model clouds for outflow model M-1 are shown in latitude and \vlsr\ in the right panel of Figure \ref{fig:M2_VLSR_vs_b}. 
Clouds in the simulation that fall below the detection limit are shown as open circles while detected clouds are colored by their value of $|z|$.
The GBT survey data are shown in the left panel.

The inclusion of detection criteria brings the simulation closer to the data, stressing the importance of observational selection effects.
The simulation at negative velocities is quite similar to the data as is the positive velocity simulation below the Galactic plane.
Above the plane, the simulation produces too many high velocities at low latitudes but the largest discrepancy is the presence of model clouds at high latitudes and high positive velocities, in an area of observational space where no \hi\ clouds are found.
We note that negative velocity clouds have been detected in GBT spectra in this area, so the effect is not instrumental.  
The discrepant clouds are all at $z > 2$ kpc.

One recurring feature evident here is that the majority of the simulated clouds have a positive LSR velocity, often by as much as $65\%$ to $35\%$.
This is expected naturally, from the larger volume within the FB that has $\cos(\beta) > 0$ than $\cos(\beta) < 0$ (Figure~\ref{fig:geometry_x-z}).
The data, however, 
show a small  excess of negative \vlsr\ clouds: 54\% to 46\%.
A result of the projection of outflow from the GC is that clouds with negative \vlsr\ are both closer to the Sun and closer to the Galactic plane than positive \vlsr\ clouds.
If we omit from  the \vlsr\ comparison clouds at latitudes $b > 10.6\dg$ (black thick boxes in Figure~\ref{fig:M2_VLSR_vs_b}), where positive values of \vlsr\ must originate at $z \gtrsim 2$ kpc, above the limit of the \hi\ cloud layer, there are virtually equal numbers of observed clouds at negative and positive \vlsr: 103 to 104.
In our simulations this equality is achieved only for the brightest clouds, suggesting that distant \hi\ clouds in the FBs are fainter than in our simulation.

\begin{figure*}
\centering
\includegraphics[width=0.95\textwidth]{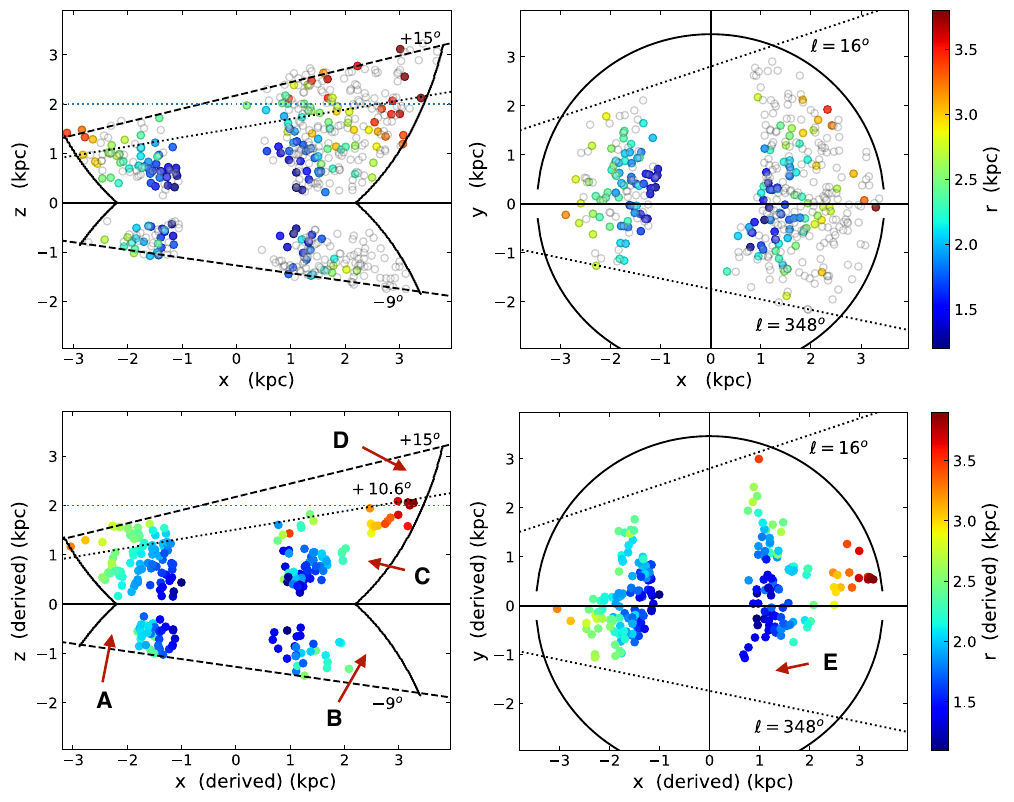}
\caption{
Simulations of the FB cloud distribution in space (upper panels), along with the data analyzed with the identical outflow model M-1 (lower panels).
In the simulations, faint grey symbols show the entire set of clouds while filled colored circles show those that that pass  detectability criteria and would appear in the GBT survey above the noise.
Colors are proportional to the derived distance from the GC.
The two left panels show projections in  $x$-$z$ coordinates and the two right panels show projections in $x$-$y$, i.e., parallel to the Galactic plane, viewed from above. 
Dashed lines show the limits of the GBT survey in latitude and dotted lines show the limitations in longitude.
Solid lines mark locations of the FB walls at $y = 0$ (left panels) and in a plane at $z = 2$ kpc (right panels).
The dotted line in the left panels at $b = 10.6\dg$ marks the upper limit to the latitude at which positive velocity \hi\ clouds are detected in this survey.
Five areas are identified in the data plots where clouds could have been detected but were not.  
Comparison with the simulations suggests that A, B, and C result from the sensitivity limits of the GBT survey, while region D requires that there be a strict upper boundary to the FB \hi\ cloud population at $z \approx 2$ kpc. 
Region E may indicate real structure in the outflow.
The empty areas in the center of all figures reflects the limit of the analysis to clouds with $|\vlsr| \geq 130$ \kms.
}
\label{fig:model-x-z-x-y}
\end{figure*}

The simulated clouds are shown in $x$-$z$ and $x$-$y$ coordinates in the top two panels of Figure~\ref{fig:model-x-z-x-y}, along with the locations of observed clouds, derived from model M-1, in the lower panels.  
It is apparent that inclusion of detectability criteria can change the results of the simulation dramatically.
There are no clouds with $x<-2$ kpc and $z<0$ kpc in the data (region marked with the letter A).  
The simulation (upper left panel) shows the same pattern, which 
arises because the GBT data at $b < 0\dg$ have a higher noise level than at $b > 0\dg$.
The simulations also show that empty regions in the data labeled B and C likely result from similar sensitivity issues to \hi\ clouds at large distances from the GC and from the Sun.

Some areas, however, appear to show real voids in the FB \hi\ population.
The most striking is the complete lack of clouds at $z > 2$ kpc (labeled D) discussed previously  (Figure~\ref{fig:M2_VLSR_vs_b}).
This results from the absence of detectable \hi\ clouds at $\vlsr > 130$ \kms\ over all $b > 10.6\dg$. 

Another deviation of the simulation from the data in the $x$-$y$ plane (right panels), i.e., when viewed from directly above the Galactic plane, is an empty region in the data labeled E.
This results from the absence of detected clouds in the area roughly bound by $348\dg \leq \ell \leq 355\dg$ and $2\dg \leq b \leq 5\dg$ (see Figure~\ref{fig:clouds_l-b-VLSR}).  
This may indicate small-scale spatial structure in the neutral outflow.

To summarize the main results from the simulations:

\begin{enumerate}
\item Many aspects of the data can be understood as arising from a  uniform population of \hi\ clouds filling the FB volume and expanding outward from the GC in an accelerating outflow. 

\item A kinematic model with outflow velocity $\Vr = 125 \ r \ \kms\ {\rm kpc^{-1}} \leq 500 \ \kms$ (M-1) gives a reasonable match to the data in latitude and \vlsr.
In agreement with the findings of  \citet{Lockman2020}, the data are not consistent with outflow of \hi\ clouds at a constant velocity.

\item Some of the features of the data appear to arise from observational selection effects caused by a) a systematic decrease in cloud  21cm line brightness  with distance from the GC; b) some dependence of cloud detectability on distance from the Sun, and c) variation in the survey noise level over the observed region.

\item There are two areas of the FBs that lack detectable \hi\ clouds, areas that do not appear naturally in the simulations.
The most striking is a complete absence of clouds at $z \gtrsim 2$ kpc.
The other is a region centered at $(\ell,b) = (357.5\dg, +4\dg)$ (region E in Figure~\ref{fig:model-x-z-x-y}).

\end{enumerate}

%%%%%%%%%%%%%%%%%%%%%%%%%%%%%%%%%%%%%
%%%%%%%%%%%%%%%%%%%%%%%%%%%%%%%%%%%%%% Longitude-velocity

\begin{figure*}[htb!]
\includegraphics[width=0.33\textwidth]{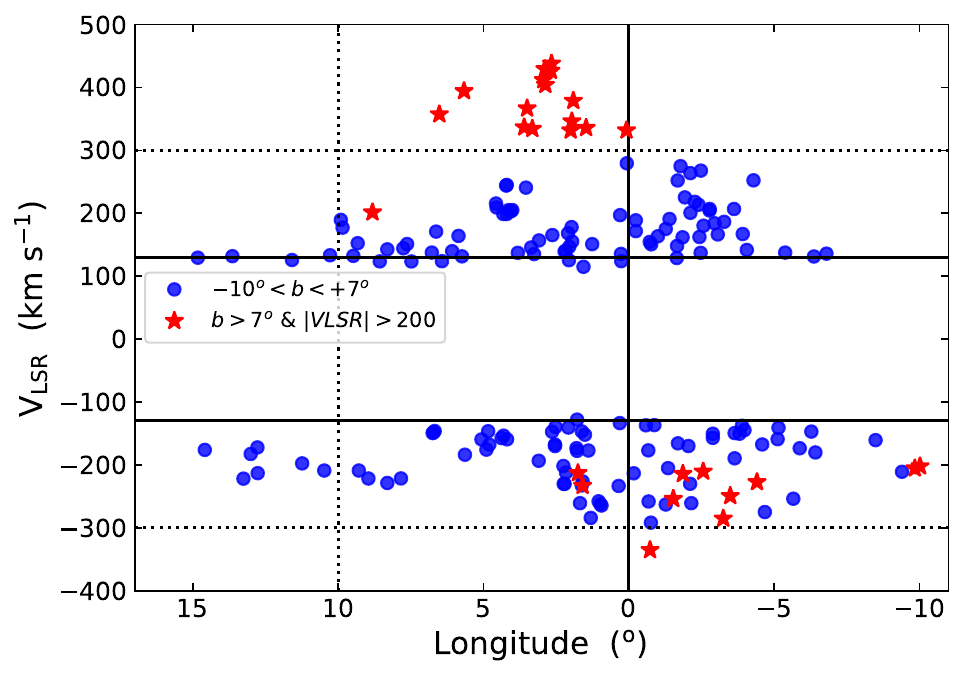}\includegraphics[width=0.33\textwidth]{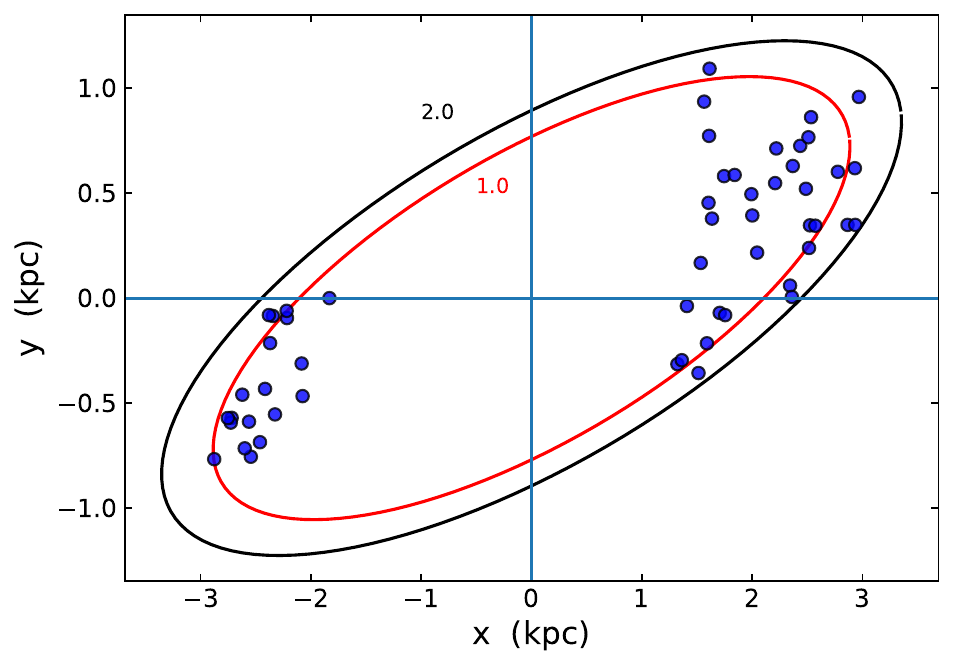}\includegraphics[width=0.33\textwidth]{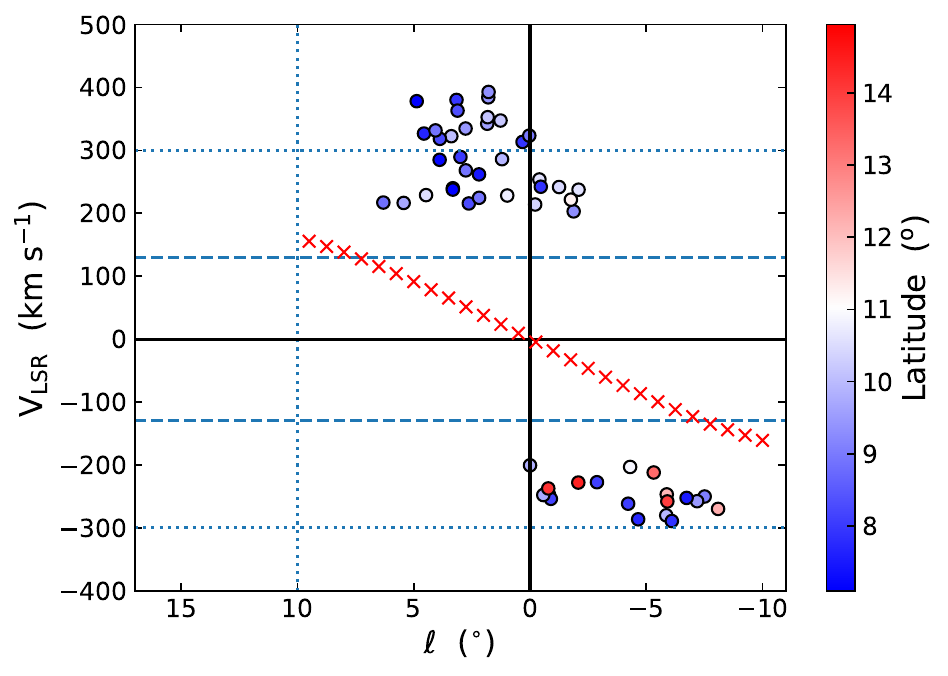}
\caption{
{\it(Left panel):}
The GBT data in longitude-velocity coordinates.
The blue circles show all  \hi\ clouds at $b \leq 7\dg$: there is no strong kinematic deviation from symmetry about $\ell = 0\dg$ for these clouds.
The red stars mark a  set of \hi\ clouds with $b > 7\dg$ and $|\vlsr| \geq 200$ \kms\  that shows a strong asymmetry with longitude: those with $\vlsr > +200$ \kms\ are located predominantly at $\ell > 0\dg$, and those with $\vlsr < -200$ \kms\ are found mainly (though not not exclusively) at $\ell < 0\dg$.
{\it Center panel:}
Simulated clouds in the M-1 model that have $b > 7\dg$, $|\vlsr| \geq 200$ \kms, and $z < 2$ kpc, and are confined to an elliptical area in $x$-$y$ within the FBs, outlined by the red and blue curves at $z = 1.0$ and 2.0 kpc, respectively.
The large empty region at low $x$ and $y$ reflects the fundamental limitation to clouds with $|\vlsr| \geq 130$ \kms.
{\it Right panel:}
This shows clouds in longitude and velocity selected from the standard model to lie within the boundaries shown in the center panel plotted in  longitude and velocity. 
These clouds follow the M-1 outflow model with no galactic rotation, and  their values of \Vr\ project to a \vlsr\ with properties similar to the high-velocity, high-latitude sample of observed clouds (red stars in the left panel).   
The red crosses show an example of what is expected for galactic rotation in this area.
The kinematics of the high-velocity, high-latitude clouds more likely derives from  spatial structure in the outflow than from a rotational component to their velocity.
}
\label{fig:long-vel}
\end{figure*}

\section{Deviations from azimuthal symmetry}
\label{sec:long-vel}

On the sky, and in the MB16 model, the FBs have a general azimuthal symmetry; their principal structural variations are vertically out of the Galaxy (Figure~\ref{fig:geometry_x-z}).
Throughout this investigation we have assumed azimuthal symmetry in the vertical structure of the FB outflow.
Unlike vertical structure, which is manifest relative to the Galactic plane, any azimuthal structure will appear as variations in cloud properties with Galactic longitude. 

Figure \ref{fig:long-vel} shows subsets of the data in longitude and \vlsr\ grouped by latitude and velocity.
The left panel shows clouds that are found at $b \leq 7\dg$ as blue circles,   about $70\%$ of the clouds in the survey.
There is no strong dependence of \vlsr\ on longitude for this set, consistent with azimuthal symmetry and an absence of  Galactic rotation in the \hi\ cloud kinematics.
In contrast, the red stars show data for clouds with  $b > 7\dg$ and $|\vlsr| > 200$ \kms.
In this high-velocity, high-latitude sample, there is a strong asymmetry: high positive \vlsr\ clouds are mainly at positive longitudes, while the high negative \vlsr\ clouds are mainly at negative latitudes.

While the kinematics of these clouds has some of the qualities expected from Galactic rotation, this possibility has significant inconsistencies: 

\begin{itemize}
    \item  Galactic rotation produces a strong systematic trend  in $\vlsr$ proportional to $\sin(\ell)$ that does not appear in the data.
    Instead, LSR velocities of this set of GBT \hi\ clouds are nearly constant with longitude. 

    \item The average \vlsr\ of the positive velocity clouds has a much greater magnitude than that of the negative velocity clouds, +364 \kms\ compared to -239 \kms, whereas from rotation we would expect near symmetry.

    \item The LSR velocities of the clouds do not have symmetry about $\ell = 0\dg$.
    
\end{itemize}

Rather than indicating rotation, these features arise naturally from an outflow with exactly the same kinematic properties as in our other models, but with a distribution of clouds that is not  azimuthally symmetric, but is elongated at an angle from the Sun-center line.

The center panel of 
Figure~\ref{fig:long-vel} shows the location of a sample set of clouds taken from 
one simulation that would produce such a longitudinal asymmetry from pure outflow.
These were selected from our standard model to have 
$b > 7\dg$ and $|\vlsr| > 200$ \kms\ but to lie within an elliptical area in $x$-$y$, whose semi-major axis is the wall of the FB ($\sim4$ kpc in length), but whose minor axis is reduced to 1 kpc.
The major axis is rotated in azimuth to $\theta = 15\dg$.

The right panel of Figure~\ref{fig:long-vel}  shows these clouds in  longitude and velocity.
Their kinematics were derived from outflow model M-1 with standard parameters of detectability, but the asymmetry in the location of the clouds projects to an asymmetry in $\vlsr(\ell)$.
For comparison, the red crosses show expectations for galactic rotation in this region.
This is for the \citet{Sofue2025}  rotation curve   evaluated at one-half the FB radius observed at a latitude of $10\dg$.
As noted above, expectations from Galactic rotation  differ in many respects from the values of \vlsr\ for the FB clouds.

This simulation does not attempt to match all features of the GBT clouds, but is sufficient to suggest that we are seeing major structural asymmetries in the large-scale outflow of neutral gas from the GC.
These asymmetries are not apparent in the  clouds at lower latitudes ($70\%$ of the sample), but appear only in clouds at some distance from the GC.
From outflow model M-1 the average distance from the GC of the clouds in the asymmetric structure is 2.7 kpc.

\section{Summary and discussion}
\label{sec:summary}

The set of 228 \hi\ clouds from the GBT surveys that are studied here  have revealed some major properties of neutral gas in the Fermi bubbles, 
some of which  were established in previous studies of more limited data sets \citep{McClure-Griffiths+13,DiTeodoro18,Lockman2020}.
The \hi\ clouds reach LSR velocities greater than any previously reported for gas associated with the Milky Way disk, and it is likely that some have a space velocity $\gtrsim 500$ \kms.

The clouds have velocities that are reasonably well modeled by a radially accelerating outflow whose magnitude increases linearly with distance from the GC: $\Vr = 125 \ r$ \kms\ ${\rm kpc^{-1}}$ up to a velocity of 500 \kms\ at $r = 4$ kpc.
Analyzed with this model for the outflow,  the \hi\ clouds are found to a distance from the GC of $r\lesssim4$ kpc.
Because our analysis is limited to clouds with $|\vlsr| \geq 130$ \kms, we have little information on the outflow at $r \lesssim 1.5$ kpc.
In the area of the GBT surveys the kinematics of absorption lines from highly ionized species such as \ion{C}{4}, \ion{Si}{4} and \ion{N}{5}
 cannot be distinguished from that of \hi. 

Several factors indicate that clouds are evolving as they flow outward in the nuclear wind.
The brightness temperature and peak \hi\ column density of the clouds show a systematic decrease with distance from the GC (Fig.~\ref{fig:TL_hists}).
The clouds also become more uniform with $r$:  
the rms scatter about the mean \TL\ decreases by an order of magnitude between regions closest to the GC and those farthest from the GC.

A major characteristic of the neutral outflow is the complete lack of detectable positive \vlsr\ \hi\ clouds over an area  $>100$ sq-deg at  $b > 10.6\dg$, corresponding to 
 an upper limit of $z \lesssim 2$ kpc from the Galactic plane for clouds within the FBs.  
Although positive \vlsr\ absorption is seen in ionized species  at a  slightly higher latitude \citep[$b = +11.2\dg$, $\vlsr = +250 \ \kms$,][]{Fox+15}, in our outflow model M-1 this still corresponds to $z \leq 2.0$ kpc (see Figure \ref{fig:TL_rho}, right panel).
The neutral cloud boundary appears to be a function of distance from the Galactic plane, not distance from the GC.  
It does not arise from sensitivity limits in the GBT data.

The simulations reproduce the main properties of the observed clouds up to $z = 2$ kpc reasonably well with three exceptions.
The first is the apparent absence of \hi\ clouds in the region around $( \ell,b) = (357.5\dg,+4\dg)$.  
This may indicate some true structure in the neutral gas distribution (Figure~\ref{fig:model-x-z-x-y}, region E).

The second discrepancy is  that the geometry of a nuclear outflow  
leads us to expect many more clouds at positive \vlsr\ than negative, whereas, when the limit of the cloud population to $z < 2$ kpc is included, there are essentially equal numbers of clouds detected at $\vlsr < 0$ as at $\vlsr > 0$.
The discrepancy likely originates in some failure of our simulations to account properly for the decrease in brightness of clouds at large $r$ and \dsun.
Whatever the source, we expect that there should be  many positive velocity \hi\ clouds lying just below the detection limit of the current survey,  
which would be consistent with an extrapolation of the observed $n(\nhi)$ distribution to lower \nhi\ (Figure~\ref{fig:properties}).

The third discrepancy (left panel of Figure~\ref{fig:long-vel}) forces us to abandon our assumption of an azimuthally symmetric outflow for the entire population of clouds.
While azimuthal symmetry seems appropriate for at least $70\%$  of the clouds -- those closest to the Galactic plane --  
asymmetries in the longitude-velocity distribution of the highest velocity clouds at $b > 7\dg$ suggest an elliptical shape to the outflow, beginning about 2.5 kpc from the GC. 
The segregation of positive \vlsr\ clouds to $\ell \gtrsim 0\dg$ and negative \vlsr\ clouds to $\ell \lesssim 0\dg$ cannot be the complete picture, however, for there are both positive and negative velocity absorption components detected in highly-ionized species  toward the AGN PDS 456 at $(\ell,b) = (10.4\dg, +11.2\dg)$ \citep{Fox+15}.

The vertical limit of \hi\ clouds to $z \lesssim 2$ kpc corresponds to a latitude limit of $22.5\dg$ at the near side 
of the FB, so we expect that negative \vlsr\ clouds might be detectable  far above the $b = 15\dg$ latitude extent of the current survey.
The negative \vlsr\ \hi\ clouds found recently  by \citet{Bordoloi2025} in  GBT observations of a field around $(\ell,b) = (+4\dg, +28\dg)$ are  plausibly part of the same population as the FB clouds in the current sample.
If these clouds are confined to the MB16 FB volume, they are located at $z \approx 2.7$ kpc,  well beyond the limit we find for positive \vlsr\ clouds.
The most negative velocity clouds in \citet{Bordoloi2025} have $\vlsr \leq -170$ \kms, which implies an outflow velocity $\Vr \geq 380$ \kms\ at their latitude.
Note that the tangent point of a radial outflow intersects  the near wall of the MB16 FB model around $b \approx 42\dg$, so a radial outflow within that volume cannot have $ \vlsr < 0$  above this latitude.

What is the ultimate fate of the \hi\ clouds in the FBs? 
In our simulation, clouds are not treated as ballistic particles but are entrained in a hot, fast wind that has lifted and accelerated them to their current location and velocity.
%and will determine their future through further acceleration and disruption.
The interaction with the hot phase will drive their future evolution: the decline in $\nhi$ with increasing $r$, together with the absence of detections above $z\simeq2$ kpc, suggests that clouds are being disrupted as they are carried outward by the wind, with their neutral hydrogen progressively converted into ionized gas. This ionized material may contribute to the diffuse ionized phase observed in absorption at higher latitudes \citep[e.g.,][]{Bordoloi+17}. Such a scenario has long been predicted by high-resolution hydrodynamical simulations of galactic winds \citep[e.g.,][]{Scannapieco&Bruggen15,Schneider+20,Gronke+20}, and is further supported by recent observations of molecular gas dissociation in several of these \hi\ clouds \citep{Noon2023,Noon+2026,DiTeodoro2026a}.

This analysis raises a number of questions, especially about the southern FB, which is relatively undersampled in our data. 
In particular, does the southern FB show a cutoff of \hi\ clouds around $ z = -2$ kpc, symmetric to that in the north?
Is there evidence of azimuthal asymmetries at $b < -7\dg$ as in the north?
These questions can only be answered with new \hi\ data covering the southern FB with good sensitivity and sky coverage.

Sensitive measurements of 21cm \hi\ line using the GBT at $9.1'$ angular resolution (22 pc at the GC) have given a detailed picture of aspects of the vertical distribution of the neutral ISM within the Milky Way's  nuclear wind that would not be possible in other galaxies.
As galactic winds are observed over a wide range of redshift these results are relevant to understanding a process that is ubiquitous throughout the Universe.

\begin{acknowledgments}
EDT was supported by the European Research Council (ERC) under grant agreement no.\ 10104075.
The new GBT data reported here were obtained under proposal codes 20B\_200, 21A\_240, and 22A\_339.  This paper also uses archival GBT data from proposals 14A\_302, 14B\_076, 14B\_461, 15B\_139, 16B\_419, 18B\_165 and 19A\_337.
The Green Bank Telescope is part of the National Radio Astronomy Observatory, a facility of the National Science Foundation operated by Associated Universities, Inc.
SC was funded in part by the U.S. National Science Foundation REU award 1852401 to NRAO/GBO.
FJL has benefited greatly from many years of discussions with W.B.~Burton on matters pertaining to neutral hydrogen in the Milky Way.
\end{acknowledgments}

\facilities{GBT}

\software{gbtgridder, GBTIDL, CASA, CARTA}

\bibliography{kinematics2025.bib}

@ARTICLE{Wang+2026,
       author = {{Wang}, Kuo-Song and {Simmonds}, Robert and {Comrie}, Angus and {Hwang}, Yu-Hsuan and {Pi{\'n}ska}, Adrianna and {Harris}, Pamela and {Moraghan}, Anthony and {Pang}, Qi and {Raul-Omar}, Carli and {Hou}, Kuan-Chou and {Aikema}, David and {Chiang}, Cheng-Chin and {Chang}, Tien-Hao and {Hsu}, Shou-Chieh and {Lin}, Ming-Yi and {Gao}, Zhen-Kai and {Huang}, Po-Sheng and {Hibbard}, John and {Ott}, Juergen and {Collier}, Jordan and {Stoehr}, Felix and {Raba}, Ryan and {Kirkham}, Kechil and {Rosolowsky}, Erik and {Kern}, Jeff and {Lee}, Chin-Fei and {Taylor}, Russ and {CARTA Collaboration}},
        title = "{CARTA{\textemdash}Cube Analysis and Rendering Tool for Astronomy: A Tool for Big Imaging Data}",
      journal = {PASP},
         year = 2026,
        month = feb,
       volume = {138},
       number = {2},
          eid = {024506},
        pages = {024506},
          doi = {10.1088/1538-3873/ae3eb4},
       adsurl = {https://ui.adsabs.harvard.edu/abs/2026PASP..138b4506W}
}

@ARTICLE{Veilleux+05,
   author = {{Veilleux}, S. and {Cecil}, G. and {Bland-Hawthorn}, J.},
    title = "{Galactic Winds}",
  journal = {\araa},
   eprint = {astro-ph/0504435},
     year = 2005,
    month = sep,
   volume = 43,
    pages = {769-826},
      doi = {10.1146/annurev.astro.43.072103.150610},
   adsurl = {http://adsabs.harvard.edu/abs/2005ARA%26A..43..769V}
}

@ARTICLE{McClure-Griffiths+13,
   author = {{McClure-Griffiths}, N.~M. and {Green}, J.~A. and {Hill}, A.~S. and 
        {Lockman}, F.~J. and {Dickey}, J.~M. and {Gaensler}, B.~M. and 
        {Green}, A.~J.},
    title = "{Atomic Hydrogen in a Galactic Center Outflow}",
  journal = {\apjl},
archivePrefix = "arXiv",
   eprint = {1304.7538},
     year = 2013,
    month = jun,
   volume = 770,
      eid = {L4},
    pages = {L4},
      doi = {10.1088/2041-8205/770/1/L4},
   adsurl = {http://adsabs.harvard.edu/abs/2013ApJ...770L...4M}
}

@ARTICLE{Scannapieco&Bruggen15,
   author = {{Scannapieco}, E. and {Br{\"u}ggen}, M.},
    title = "{The Launching of Cold Clouds by Galaxy Outflows. I. Hydrodynamic Interactions with Radiative Cooling}",
  journal = {\apj},
archivePrefix = "arXiv",
   eprint = {1503.06800},
     year = 2015,
    month = jun,
   volume = 805,
      eid = {158},
    pages = {158},
      doi = {10.1088/0004-637X/805/2/158},
   adsurl = {http://adsabs.harvard.edu/abs/2015ApJ...805..158S}
}

@ARTICLE{Su+10,
   author = {{Su}, M. and {Slatyer}, T.~R. and {Finkbeiner}, D.~P.},
    title = "{Giant Gamma-ray Bubbles from Fermi-LAT: Active Galactic Nucleus Activity or Bipolar Galactic Wind?}",
  journal = {\apj},
archivePrefix = "arXiv",
   eprint = {1005.5480},
 primaryClass = "astro-ph.HE",
     year = 2010,
    month = dec,
   volume = 724,
    pages = {1044-1082},
      doi = {10.1088/0004-637X/724/2/1044},
   adsurl = {http://adsabs.harvard.edu/abs/2010ApJ...724.1044S}
}

@ARTICLE{Bland-Hawthorn_Cohen03,
   author = {{Bland-Hawthorn}, J. and {Cohen}, M.},
    title = "{The Large-Scale Bipolar Wind in the Galactic Center}",
  journal = {\apj},
   eprint = {astro-ph/0208553},
     year = 2003,
    month = jan,
   volume = 582,
    pages = {246-256},
      doi = {10.1086/344573},
   adsurl = {http://adsabs.harvard.edu/abs/2003ApJ...582..246B}
}

@ARTICLE{Fox+15,
   author = {{Fox}, A.~J. and {Bordoloi}, R. and {Sav
 age}, B.~D. and {Lockman}, F.~J. and 
        {Jenkins}, E.~B. and {Wakker}, B.~P. and {Bland-Hawthorn}, J. and 
        {Hernandez}, S. and {Kim}, T.-S. and {Benjamin}, R.~A. and {Bowen}, D.~V. and 
        {Tumlinson}, J.},
    title = "{Probing the Fermi Bubbles in Ultraviolet Absorption: A Spectroscopic Signature of the Milky Way's Biconical Nuclear Outflow}",
  journal = {\apjl},
archivePrefix = "arXiv",
   eprint = {1412.1480},
     year = 2015,
    month = jan,
   volume = 799,
      eid = {L7},
    pages = {L7},
      doi = {10.1088/2041-8205/799/1/L7},
   adsurl = {http://adsabs.harvard.edu/abs/2015ApJ...799L...7F}
}

@ARTICLE{Bordoloi+17,
   author = {{Bordoloi}, R. and {Fox}, A.~J. and {Lockman}, F.~J. and {Wakker}, B.~P. and 
        {Jenkins}, E.~B. and {Savage}, B.~D. and {Hernandez}, S. and 
        {Tumlinson}, J. and {Bland-Hawthorn}, J. and {Kim}, T.-S.},
    title = "{Mapping the Nuclear Outflow of the Milky Way: Studying the Kinematics and Spatial Extent of the Northern Fermi Bubble}",
  journal = {\apj},
archivePrefix = "arXiv",
   eprint = {1612.01578},
     year = 2017,
    month = jan,
   volume = 834,
      eid = {191},
    pages = {191},
      doi = {10.3847/1538-4357/834/2/191},
   adsurl = {http://adsabs.harvard.edu/abs/2017ApJ...834..191B}
}

@ARTICLE{Lockman84,
   author = {{Lockman}, F.~J.},
    title = "{The H I halo in the inner galaxy}",
  journal = {\apj},
     year = 1984,
    month = aug,
   volume = 283,
    pages = {90-97},
      doi = {10.1086/162277},
   adsurl = {http://adsabs.harvard.edu/abs/1984ApJ...283...90L}
}

@ARTICLE{Lockman_McClureGriffiths16,  
   author = {{Lockman}, F.~J. and {McClure-Griffiths}, N.~M.},
    title = "{Tracing the Milky Way Nuclear Wind with 21cm Atomic Hydrogen Emission}",
  journal = {\apj},
archivePrefix = "arXiv",
   eprint = {1605.01140},
     year = 2016,
    month = aug,
   volume = 826,
      eid = {215},
    pages = {215},
      doi = {10.3847/0004-637X/826/2/215},
   adsurl = {http://adsabs.harvard.edu/abs/2016ApJ...826..215L}
}

@ARTICLE{Boothroyd2011,
  author = {{Boothroyd}, A.~I. and {Blagrave}, K. and {Lockman}, F.~J. and
        {Martin}, P.~G. and {Pinheiro Gon{\c c}alves}, D. and {Srikanth}, S.
        },
   title = "{Accurate galactic 21-cm H I measurements with the NRAO Green
Bank Telescope}",
 journal = {\aap},
archivePrefix = "arXiv",
  eprint = {1110.1765},
primaryClass = "astro-ph.IM",
    year = 2011,
   month = dec,
  volume = 536,
     eid = {A81},
   pages = {A81},
     doi = {10.1051/0004-6361/201117656},
  adsurl = {http://adsabs.harvard.edu/abs/2011A%26A...536A..81B}
}

@ARTICLE{Lockman2020,
       author = {{Lockman}, Felix J. and {Di Teodoro}, Enrico M. and {McClure-Griffiths}, N.~M.},
        title = "{Observation of Acceleration of H I Clouds within the Fermi Bubbles}",
      journal = {\apj},
         year = 2020,
        month = jan,
       volume = {888},
       number = {1},
          eid = {51},
        pages = {51},
          doi = {10.3847/1538-4357/ab55d8},
archivePrefix = {arXiv},
       eprint = {1911.06864},
 primaryClass = {astro-ph.GA},
       adsurl = {https://ui.adsabs.harvard.edu/abs/2020ApJ...888...51L}
}

@ARTICLE{Savage+17,
   author = {{Savage}, B.~D. and {Kim}, T.-S. and {Fox}, A.~J. and {Massa}, D. and 
        {Bordoloi}, R. and {Jenkins}, E.~B. and {Lehner}, N. and {Bland-Hawthorn}, J. and 
        {Lockman}, F.~J. and {Hernandez}, S. and {Wakker}, B.~P.},
    title = "{Probing the Outflowing Multiphase Gas {\sim}1 kpc below the Galactic Center}",
  journal = {\apjs},
archivePrefix = "arXiv",
   eprint = {1707.06942},
     year = 2017,
    month = oct,
   volume = 232,
      eid = {25},
    pages = {25},
      doi = {10.3847/1538-4365/aa8f4c},
   adsurl = {http://adsabs.harvard.edu/abs/2017ApJS..232...25S}
}

@ARTICLE{MillerBregman16,
   author = {{Miller}, M.~J. and {Bregman}, J.~N.},
    title = "{The Interaction of the Fermi Bubbles with the Milky Way{\rsquo}s Hot Gas Halo}",
  journal = {\apj},
archivePrefix = "arXiv",
   eprint = {1607.04906},
     year = 2016,
    month = sep,
   volume = 829,
      eid = {9},
    pages = {9},
      doi = {10.3847/0004-637X/829/1/9},
   adsurl = {http://adsabs.harvard.edu/abs/2016ApJ...829....9M}
}

@ARTICLE{Lockman2002,
       author = {{Lockman}, Felix J.},
        title = "{Discovery of a Population of H I Clouds in the Galactic Halo}",
      journal = {\apjl},
         year = 2002,
        month = nov,
       volume = {580},
       number = {1},
        pages = {L47-L50},
          doi = {10.1086/345495},
archivePrefix = {arXiv},
       eprint = {astro-ph/0210424},
 primaryClass = {astro-ph},
       adsurl = {https://ui.adsabs.harvard.edu/abs/2002ApJ...580L..47L}
}

@ARTICLE{Stil2006,
       author = {{Stil}, J.~M. and {Lockman}, Felix. J. and {Taylor}, A.~R. and {Dickey}, J.~M. and {Kavars}, D.~W. and {Martin}, P.~G. and {Rothwell}, T.~A. and {Boothroyd}, A.~I. and {McClure-Griffiths}, N.~M.},
        title = "{Compact H I Clouds at High Forbidden Velocities in the Inner Galaxy}",
      journal = {\apj},
         year = 2006,
        month = jan,
       volume = {637},
       number = {1},
        pages = {366-379},
          doi = {10.1086/498347},
archivePrefix = {arXiv},
       eprint = {astro-ph/0509730},
 primaryClass = {astro-ph},
       adsurl = {https://ui.adsabs.harvard.edu/abs/2006ApJ...637..366S}
}

@ARTICLE{DiTeodoro18,
       author = {{Di Teodoro}, Enrico M. and {McClure-Griffiths}, N.~M. and {Lockman}, Felix J. and {Denbo}, Sara R. and {Endsley}, Ryan and {Ford}, H. Alyson and {Harrington}, Kevin},
        title = "{Blowing in the Milky Way Wind: Neutral Hydrogen Clouds Tracing the Galactic Nuclear Outflow}",
      journal = {\apj},
         year = 2018,
        month = mar,
       volume = {855},
       number = {1},
          eid = {33},
        pages = {33},
          doi = {10.3847/1538-4357/aaad6a},
archivePrefix = {arXiv},
       eprint = {1802.02152},
 primaryClass = {astro-ph.GA},
       adsurl = {https://ui.adsabs.harvard.edu/abs/2018ApJ...855...33D}
}

@ARTICLE{DiTeodoro2020,
       author = {{Di Teodoro}, Enrico M. and {McClure-Griffiths}, N.~M. and {Lockman}, Felix J. and {Armillotta}, Lucia},
        title = "{Cold gas in the Milky Way's nuclear wind}",
      journal = {\nat},
         year = 2020,
        month = aug,
       volume = {584},
       number = {7821},
        pages = {364-367},
          doi = {10.1038/s41586-020-2595-z},
archivePrefix = {arXiv},
       eprint = {2008.09121},
 primaryClass = {astro-ph.GA},
       adsurl = {https://ui.adsabs.harvard.edu/abs/2020Natur.584..364D}
}

@ARTICLE{Noon2023,
       author = {{Noon}, Karlie A. and {Krumholz}, Mark R. and {Di Teodoro}, Enrico M. and {McClure-Griffiths}, Naomi M. and {Lockman}, Felix J. and {Armillotta}, Lucia},
        title = "{Direct observations of the atomic-molecular phase transition in the Milky Way's nuclear wind}",
      journal = {\mnras},
         year = 2023,
        month = sep,
       volume = {524},
       number = {1},
        pages = {1258-1268},
          doi = {10.1093/mnras/stad1890},
archivePrefix = {arXiv},
       eprint = {2304.06356},
 primaryClass = {astro-ph.GA},
       adsurl = {https://ui.adsabs.harvard.edu/abs/2023MNRAS.524.1258N}
}

@ARTICLE{Keeney2006,
       author = {{Keeney}, Brian A. and {Danforth}, Charles W. and {Stocke}, John T. and {Penton}, Steven V. and {Shull}, J. Michael and {Sembach}, Kenneth R.},
        title = "{Does the Milky Way Produce a Nuclear Galactic Wind?}",
      journal = {\apj},
         year = 2006,
        month = aug,
       volume = {646},
       number = {2},
        pages = {951-964},
          doi = {10.1086/505128},
archivePrefix = {arXiv},
       eprint = {astro-ph/0604323},
 primaryClass = {astro-ph},
       adsurl = {https://ui.adsabs.harvard.edu/abs/2006ApJ...646..951K}
}

@INPROCEEDINGS{Prestage2015,
  author = {{Prestage}, R.~M. and {Bloss}, M. and {Brandt}, J. and
{Chen}, H. and
	{Creager}, R. and {Demorest}, P. and {Ford}, J. and {Jones}, G. and
	{Kepley}, A. and {Kobelski}, A. and {Marganian}, P. and {Mello}, M. and
	{McMahon}, D. and {McCullough}, R. and {Ray}, J. and {Roshi}, D.~A. and
	{Werthimer}, D. and {Whitehead}, M.},
   title = "{The versatile GBT astronomical spectrometer (VEGAS): Current
status and future plans}",
booktitle = {2015 URSI-USNC Radio Science Meeting, 19-24 July 2015,
Vancouver, BC, Canada},
    year = 2015,
   month = jul,
     doi = {10.1109/USNC-URSI.2015.7303578},
  adsurl = {http://adsabs.harvard.edu/abs/2015ursi.confE...4P}
}

@ARTICLE{Prestage2009,
       author = {{Prestage}, R.~M. and {Constantikes}, K.~T. and {Hunter}, T.~R. and {King}, L.~J. and {Lacasse}, R.~J. and {Lockman}, F.~J. and {Norrod}, R.~D.},
        title = "{The Green Bank Telescope}",
      journal = {IEEE Proceedings},
         year = 2009,
        month = aug,
       volume = {97},
       number = {8},
        pages = {1382-1390},
          doi = {10.1109/JPROC.2009.2015467},
       adsurl = {https://ui.adsabs.harvard.edu/abs/2009IEEEP..97.1382P}
}

@ARTICLE{Carretti2013,
   author = {{Carretti}, E. and {Crocker}, R.~M. and {Staveley-Smith}, L. and 
	{Haverkorn}, M. and {Purcell}, C. and {Gaensler}, B.~M. and 
	{Bernardi}, G. and {Kesteven}, M.~J. and {Poppi}, S.},
    title = "{Giant magnetized outflows from the centre of the Milky Way}",
  journal = {\nat},
archivePrefix = "arXiv",
   eprint = {1301.0512},
 primaryClass = "astro-ph.GA",
     year = 2013,
    month = jan,
   volume = 493,
    pages = {66-69},
      doi = {10.1038/nature11734},
   adsurl = {http://adsabs.harvard.edu/abs/2013Natur.493...66C}
}

@ARTICLE{Burton1971,
       author = {{Burton}, W.~B.},
        title = "{Galactic Structure Derived from Neutral Hydrogen Observations Using Kinematic Models Based on the Density-wave Theory}",
      journal = {\aap},
         year = 1971,
        month = jan,
       volume = {10},
        pages = {76},
       adsurl = {https://ui.adsabs.harvard.edu/abs/1971A&A....10...76B}
}

@ARTICLE{DickeyLockman1990,
       author = {{Dickey}, John M. and {Lockman}, Felix J.},
        title = "{H I in the galaxy.}",
      journal = {\araa},
         year = 1990,
        month = jan,
       volume = {28},
        pages = {215-261},
          doi = {10.1146/annurev.aa.28.090190.001243},
       adsurl = {https://ui.adsabs.harvard.edu/abs/1990ARA&A..28..215D}
}

@ARTICLE{Ford2010,
       author = {{Ford}, H. Alyson and {Lockman}, Felix J. and {McClure-Griffiths}, N.~M.},
        title = "{Milky Way Disk-Halo Transition in H I: Properties of the Cloud Population}",
      journal = {\apj},
         year = 2010,
        month = oct,
       volume = {722},
       number = {1},
        pages = {367-379},
          doi = {10.1088/0004-637X/722/1/367},
archivePrefix = {arXiv},
       eprint = {1008.2760},
 primaryClass = {astro-ph.GA},
       adsurl = {https://ui.adsabs.harvard.edu/abs/2010ApJ...722..367F}
}

@ARTICLE{GRAVITY2021,
       author = {{GRAVITY Collaboration} and {Abuter}, R. and {Amorim}, A. and {Baub{\"o}ck}, M. and {Berger}, J.~P. and {Bonnet}, H. and {Brandner}, W. and {Cl{\'e}net}, Y. and {Davies}, R. and {de Zeeuw}, P.~T. and {Dexter}, J. and {Dallilar}, Y. and {Drescher}, A. and {Eckart}, A. and {Eisenhauer}, F. and {F{\"o}rster Schreiber}, N.~M. and {Garcia}, P. and {Gao}, F. and {Gendron}, E. and {Genzel}, R. and {Gillessen}, S. and {Habibi}, M. and {Haubois}, X. and {Hei{\ss}el}, G. and {Henning}, T. and {Hippler}, S. and {Horrobin}, M. and {Jim{\'e}nez-Rosales}, A. and {Jochum}, L. and {Jocou}, L. and {Kaufer}, A. and {Kervella}, P. and {Lacour}, S. and {Lapeyr{\`e}re}, V. and {Le Bouquin}, J. -B. and {L{\'e}na}, P. and {Lutz}, D. and {Nowak}, M. and {Ott}, T. and {Paumard}, T. and {Perraut}, K. and {Perrin}, G. and {Pfuhl}, O. and {Rabien}, S. and {Rodr{\'\i}guez-Coira}, G. and {Shangguan}, J. and {Shimizu}, T. and {Scheithauer}, S. and {Stadler}, J. and {Straub}, O. and {Straubmeier}, C. and {Sturm}, E. and {Tacconi}, L.~J. and {Vincent}, F. and {von Fellenberg}, S. and {Waisberg}, I. and {Widmann}, F. and {Wieprecht}, E. and {Wiezorrek}, E. and {Woillez}, J. and {Yazici}, S. and {Young}, A. and {Zins}, G.},
        title = "{Improved GRAVITY astrometric accuracy from modeling optical aberrations}",
      journal = {\aap},
         year = 2021,
        month = mar,
       volume = {647},
          eid = {A59},
        pages = {A59},
          doi = {10.1051/0004-6361/202040208},
archivePrefix = {arXiv},
       eprint = {2101.12098},
 primaryClass = {astro-ph.GA},
       adsurl = {https://ui.adsabs.harvard.edu/abs/2021A&A...647A..59G}
}

@ARTICLE{Weiner1999,
       author = {{Weiner}, Benjamin J. and {Sellwood}, J.~A.},
        title = "{The Properties of the Galactic Bar Implied by Gas Kinematics in the Inner Milky Way}",
      journal = {\apj},
         year = 1999,
        month = oct,
       volume = {524},
       number = {1},
        pages = {112-128},
          doi = {10.1086/307786},
archivePrefix = {arXiv},
       eprint = {astro-ph/9904130},
 primaryClass = {astro-ph},
       adsurl = {https://ui.adsabs.harvard.edu/abs/1999ApJ...524..112W}
}

@ARTICLE{Binney1991,
       author = {{Binney}, James and {Gerhard}, Ortwin E. and {Stark}, Antony A. and {Bally}, John and {Uchida}, Keven I.},
        title = "{Understanding the kinematics of Galactic Centre gas.}",
      journal = {\mnras},
         year = 1991,
        month = sep,
       volume = {252},
        pages = {210},
          doi = {10.1093/mnras/252.2.210},
       adsurl = {https://ui.adsabs.harvard.edu/abs/1991MNRAS.252..210B}
}

@ARTICLE{Sormani2015,
       author = {{Sormani}, Mattia C. and {Magorrian}, John},
        title = "{Recognizing the fingerprints of the Galactic bar: a quantitative approach to comparing model (l, v) distributions to observations}",
      journal = {\mnras},
         year = 2015,
        month = feb,
       volume = {446},
       number = {4},
        pages = {4186-4204},
          doi = {10.1093/mnras/stu2316},
archivePrefix = {arXiv},
       eprint = {1410.8073},
 primaryClass = {astro-ph.GA},
       adsurl = {https://ui.adsabs.harvard.edu/abs/2015MNRAS.446.4186S}
}

@ARTICLE{Sarkar2024,
       author = {{Sarkar}, Kartick C.},
        title = "{The Fermi/eROSITA bubbles: a look into the nuclear outflow from the Milky Way}",
      journal = {\aapr},
         year = 2024,
        month = mar,
       volume = {32},
       number = {1},
          eid = {1},
        pages = {1},
          doi = {10.1007/s00159-024-00152-1},
archivePrefix = {arXiv},
       eprint = {2403.09824},
 primaryClass = {astro-ph.HE},
       adsurl = {https://ui.adsabs.harvard.edu/abs/2024A&ARv..32....1S}
}

@ARTICLE{Afruni+2026,
       author = {{Afruni}, Andrea and {Di Teodoro}, Enrico M. and {Armillotta}, Lucia and {Lynn}, Callum A. and {McClure-Griffiths}, Naomi M.},
        title = "{Modeling the Milky Way wind: Supernova-driven outflows accelerate HI clouds near the Galactic center}",
      journal = {A\&A},
         year = 2026,
        month = feb,
       volume = {706},
          eid = {A297},
        pages = {A297},
          doi = {10.1051/0004-6361/202556813},
archivePrefix = {arXiv},
       eprint = {2601.05314},
 primaryClass = {astro-ph.GA},
       adsurl = {https://ui.adsabs.harvard.edu/abs/2026A&A...706A.297A}
}

@ARTICLE{Noon+2026,
       author = {{Noon}, Karlie A. and {Krumholz}, Mark R. and {McClure-Griffiths}, Naomi M. and {Di Teodoro}, Enrico M. and {Armillotta}, Lucia},
        title = "{Modelling the non-equilibrium chemistry of the Milky Way's cold nuclear wind}",
      journal = {MNRAS},
         year = 2026,
        month = may,
       volume = {548},
       number = {3},
          eid = {stag715},
        pages = {stag715},
          doi = {10.1093/mnras/stag715},
archivePrefix = {arXiv},
       eprint = {2602.13580},
 primaryClass = {astro-ph.GA},
       adsurl = {https://ui.adsabs.harvard.edu/abs/2026MNRAS.548ag715N}
}

@ARTICLE{Schneider+20,
       author = {{Schneider}, Evan E. and {Ostriker}, Eve C. and {Robertson}, Brant E. and
         {Thompson}, Todd A.},
        title = "{The Physical Nature of Starburst-driven Galactic Outflows}",
      journal = {\apj},
         year = 2020,
        month = may,
       volume = {895},
       number = {1},
          eid = {43},
        pages = {43},
          doi = {10.3847/1538-4357/ab8ae8},
       adsurl = {https://ui.adsabs.harvard.edu/abs/2020ApJ...895...43S}
}

@ARTICLE{Gronke+20,
       author = {{Gronke}, Max and {Oh}, S. Peng},
        title = "{How cold gas continuously entrains mass and momentum from a hot wind}",
      journal = {\mnras},
         year = 2020,
        month = feb,
       volume = {492},
       number = {2},
        pages = {1970-1990},
          doi = {10.1093/mnras/stz3332},
archivePrefix = {arXiv},
       eprint = {1907.04771},
 primaryClass = {astro-ph.GA},
       adsurl = {https://ui.adsabs.harvard.edu/abs/2020MNRAS.492.1970G}
}

@ARTICLE{DiTeodoro2026a,
       author = {{Di Teodoro}, Enrico M. and {Heyer}, Mark and {Krumholz}, Mark R. and {Armillotta}, Lucia and {Lockman}, Felix J. and {Afruni}, Andrea and {Busch}, Michael P. and {McClure-Griffiths}, N.~M. and {Noon}, Karlie A. and {Peschken}, Nicolas and {Yu}, Qingzheng},
        title = "{A survey of molecular clouds in the Galactic center's outflow}",
      journal = {arXiv e-prints},
         year = 2026,
        month = jan,
          eid = {arXiv:2601.07907},
        pages = {arXiv:2601.07907},
          doi = {10.48550/arXiv.2601.07907},
archivePrefix = {arXiv},
       eprint = {2601.07907},
 primaryClass = {astro-ph.GA},
       adsurl = {https://ui.adsabs.harvard.edu/abs/2026arXiv260107907D}
}

@ARTICLE{SuZhang2022,
       author = {{Su}, Yang and {Zhang}, Shiyu and {Yang}, Ji and {Yan}, Qing-Zeng and {Sun}, Yan and {Wang}, Hongchi and {Zhang}, Shaobo and {Chen}, Xuepeng and {Chen}, Zhiwei and {Zhou}, Xin and {Yuan}, Lixia},
        title = "{CO Emission Delineating the Interface between the Milky Way Nuclear Wind Cavity and the Gaseous Disk}",
      journal = {\apj},
         year = 2022,
        month = may,
       volume = {930},
       number = {2},
          eid = {112},
        pages = {112},
          doi = {10.3847/1538-4357/ac63b3},
archivePrefix = {arXiv},
       eprint = {2204.00728},
 primaryClass = {astro-ph.GA},
       adsurl = {https://ui.adsabs.harvard.edu/abs/2022ApJ...930..112S}
}

@ARTICLE{Ashley2022,
       author = {{Ashley}, Trisha and {Fox}, Andrew J. and {Cashman}, Frances H. and {Lockman}, Felix J. and {Bordoloi}, Rongmon and {Jenkins}, Edward B. and {Wakker}, Bart P. and {Karim}, Tanveer},
        title = "{Diverse metallicities of Fermi bubble clouds indicate dual origins in the disk and halo}",
      journal = {Nature Astronomy},
         year = 2022,
        month = jul,
       volume = {6},
        pages = {968-975},
          doi = {10.1038/s41550-022-01720-0},
archivePrefix = {arXiv},
       eprint = {2207.08838},
 primaryClass = {astro-ph.GA},
       adsurl = {https://ui.adsabs.harvard.edu/abs/2022NatAs...6..968A}
}

@ARTICLE{Sofue2025,
       author = {{Sofue}, Yoshiaki and {Kohno}, Mikito},
        title = "{The inner rotation curve of the Milky Way}",
      journal = {\pasj},
         year = 2025,
        month = dec,
       volume = {77},
       number = {6},
        pages = {1335-1349},
          doi = {10.1093/pasj/psaf114},
archivePrefix = {arXiv},
       eprint = {2509.23581},
 primaryClass = {astro-ph.GA},
       adsurl = {https://ui.adsabs.harvard.edu/abs/2025PASJ...77.1335S}
}

@ARTICLE{Predehl2020,
       author = {{Predehl}, P. and {Sunyaev}, R.~A. and {Becker}, W. and {Brunner}, H. and {Burenin}, R. and {Bykov}, A. and {Cherepashchuk}, A. and {Chugai}, N. and {Churazov}, E. and {Doroshenko}, V. and {Eismont}, N. and {Freyberg}, M. and {Gilfanov}, M. and {Haberl}, F. and {Khabibullin}, I. and {Krivonos}, R. and {Maitra}, C. and {Medvedev}, P. and {Merloni}, A. and {Nandra}, K. and {Nazarov}, V. and {Pavlinsky}, M. and {Ponti}, G. and {Sanders}, J.~S. and {Sasaki}, M. and {Sazonov}, S. and {Strong}, A.~W. and {Wilms}, J.},
        title = "{Detection of large-scale X-ray bubbles in the Milky Way halo}",
      journal = {\nat},
         year = 2020,
        month = dec,
       volume = {588},
       number = {7837},
        pages = {227-231},
          doi = {10.1038/s41586-020-2979-0},
archivePrefix = {arXiv},
       eprint = {2012.05840},
 primaryClass = {astro-ph.GA},
       adsurl = {https://ui.adsabs.harvard.edu/abs/2020Natur.588..227P}
}

@ARTICLE{Sofue2021,
       author = {{Sofue}, Yoshiaki and {Kataoka}, Jun},
        title = "{Interaction of the galactic-centre super bubbles with the gaseous disc}",
      journal = {\mnras},
         year = 2021,
        month = sep,
       volume = {506},
       number = {2},
        pages = {2170-2180},
          doi = {10.1093/mnras/stab1857},
archivePrefix = {arXiv},
       eprint = {2106.14955},
 primaryClass = {astro-ph.GA},
       adsurl = {https://ui.adsabs.harvard.edu/abs/2021MNRAS.506.2170S}
}

@ARTICLE{Bordoloi2025,
       author = {{Bordoloi}, Rongmon and {Fox}, Andrew J. and {Lockman}, Felix J.},
        title = "{A New High-latitude H I Cloud Complex Entrained in the Northern Fermi Bubble}",
      journal = {\apjl},
         year = 2025,
        month = jul,
       volume = {987},
       number = {2},
          eid = {L32},
        pages = {L32},
          doi = {10.3847/2041-8213/addd16},
archivePrefix = {arXiv},
       eprint = {2504.21091},
 primaryClass = {astro-ph.GA},
       adsurl = {https://ui.adsabs.harvard.edu/abs/2025ApJ...987L..32B}
}

@ARTICLE{Tumlinson2017,
       author = {{Tumlinson}, Jason and {Peeples}, Molly S. and {Werk}, Jessica K.},
        title = "{The Circumgalactic Medium}",
      journal = {\araa},
         year = 2017,
        month = aug,
       volume = {55},
       number = {1},
        pages = {389-432},
          doi = {10.1146/annurev-astro-091916-055240},
archivePrefix = {arXiv},
       eprint = {1709.09180},
 primaryClass = {astro-ph.GA},
       adsurl = {https://ui.adsabs.harvard.edu/abs/2017ARA&A..55..389T}
}

@ARTICLE{Veilleux2020,
       author = {{Veilleux}, Sylvain and {Maiolino}, Roberto and {Bolatto}, Alberto D. and {Aalto}, Susanne},
        title = "{Cool outflows in galaxies and their implications}",
      journal = {\aapr},
         year = 2020,
        month = apr,
       volume = {28},
       number = {1},
          eid = {2},
        pages = {2},
          doi = {10.1007/s00159-019-0121-9},
archivePrefix = {arXiv},
       eprint = {2002.07765},
 primaryClass = {astro-ph.GA},
       adsurl = {https://ui.adsabs.harvard.edu/abs/2020A&ARv..28....2V}
}

@ARTICLE{Heckman2017,
       author = {{Heckman}, Timothy M. and {Thompson}, Todd A.},
        title = "{Galactic Winds and the Role Played by Massive Stars}",
      journal = {arXiv e-prints},
         year = 2017,
        month = jan,
          eid = {arXiv:1701.09062},
        pages = {arXiv:1701.09062},
          doi = {10.48550/arXiv.1701.09062},
archivePrefix = {arXiv},
       eprint = {1701.09062},
 primaryClass = {astro-ph.GA},
       adsurl = {https://ui.adsabs.harvard.edu/abs/2017arXiv170109062H}
}
\bibliographystyle{aasjournal}



\appendix   

\section{Detectability of clouds in the GBT survey}
\label{app:TL-vs-dSun}

In order to study observational selection effects including the detectability of \hi\ clouds at different distances, we select a sample of clouds from 
the area bound by $|\ell| \leq 5\dg$ and $5\dg \leq \ b \leq 15\dg$, which we call the "deep" region.
It is covered in the survey at a noise level that is both low and relatively uniform across the region.
From the geometry of the FB it is clear that clouds in the deep region with $\vlsr < 0$, i.e., those on the near side of the FBs, must come from a relatively small range of \dsun.
Using the outflow model M-1 defined in Section \ref{sec:Vout_limits}, the 39 clouds in the "deep, near" region have a derived mean $\dsun = 6.3\pm0.4$ kpc.

The distribution of the median \TL\ for these clouds is shown in Figure~\ref{fig:Deep_dist}, along with an exponential fit  that has a scale of 0.13 K.
In simulating the observations we will assume that this represents the intrinsic distribution of \hi\ cloud values of \TL\ when the sample is observed with the GBT at a distance of 6.3 kpc.

The detectability of an \hi\ cloud is proportional to the line brightness, \TL, which we measured over a $10'\times10'$ area to derive the basic cloud properties used in this investigation.
The median diameter of GBT clouds in the \citet{DiTeodoro18} sample is $\sim 35'$, so the distance-dependence of $\TL$ is unlikely to follow a simple inverse-square law.
It can, however, be estimated from the data itself by integrating over rectangles of different areas. 
For example, the average of the data over a $20'\times 20'$ rectangle  gives the cloud's \TL\ as it would be measured were it at twice its actual distance.

Over the area of the GBT survey, the path through the FB never exceeds a factor of 2.4 in distance from the Sun, and is more typically a factor $\leq2$  (see Figure~\ref{fig:geometry_x-z}).
Thus the difference between line brightness measured over the $10'$ and the $20'$ areas should cover the most important  distance-dependent effects.   
The results are shown in Figure \ref{fig:TL_scaled} for the faintest $75\%$ of the clouds.
The brighter clouds follow the same pattern.

Any FB \hi\ clouds that have angular size  $\leq10'$ would have their $20'$ measurements lie on the dashed  curve appropriate for an inverse-square relationship.  
We see, instead, evidence of a much more gradual distance dependence, consistent with the fact that most clouds are extended to the $9.1'$ GBT antenna beam  \citep{DiTeodoro18}.
From these measurements we derive an empirical relationship $\TL \propto 6.3 / \dsun$ with \dsun\ in kpc, where the factor 6.3 is the mean distance to the deep-near sample derived from outflow  model M-1.
While this distance-dependence is fairly mild, Figure~\ref{fig:TL_scaled} suggests that about half of the 39 clouds in the deep-near sample would not be detectable at the greatest  distances within the FB volume covered by the GBT survey.
This is taken into account in simulations of the data in Section \ref{sec:models}.

\begin{figure}[!b]
\centering
\includegraphics[width=0.55\textwidth]{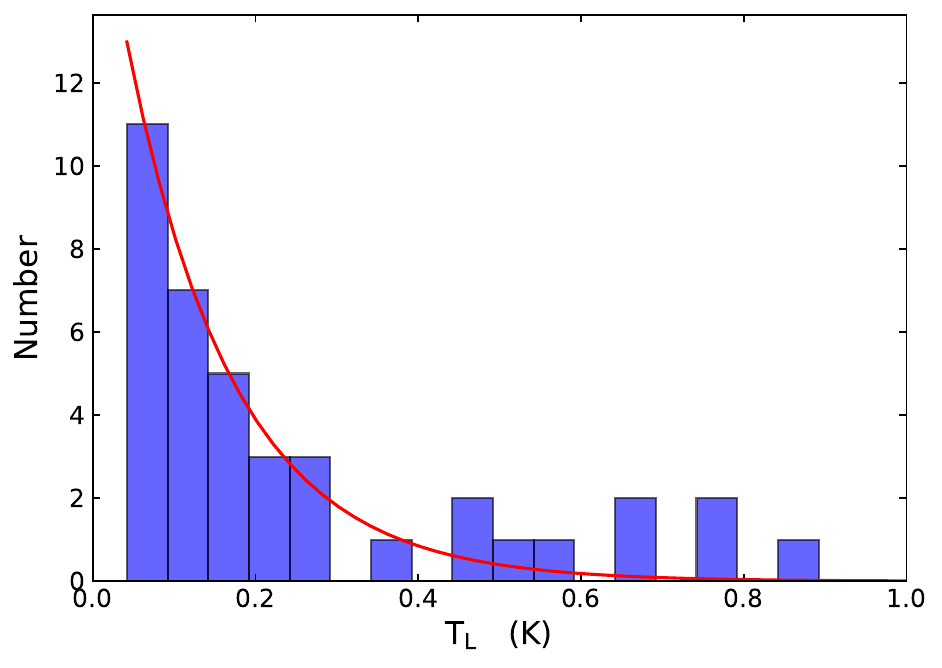}
\caption{
The  21cm \hi\ line brightness temperature, \TL, for the 39  clouds in the deep-near sample. 
We adopt this as the intrinsic distribution of \TL\ to be used in simulations of the detectability of clouds at different locations.
The curve shows the fit of an exponential to the binned values.  It has a scale of 0.13 K.
}
\label{fig:Deep_dist}
\end{figure}

\begin{figure}
\centering
\includegraphics[width=0.55\textwidth]
{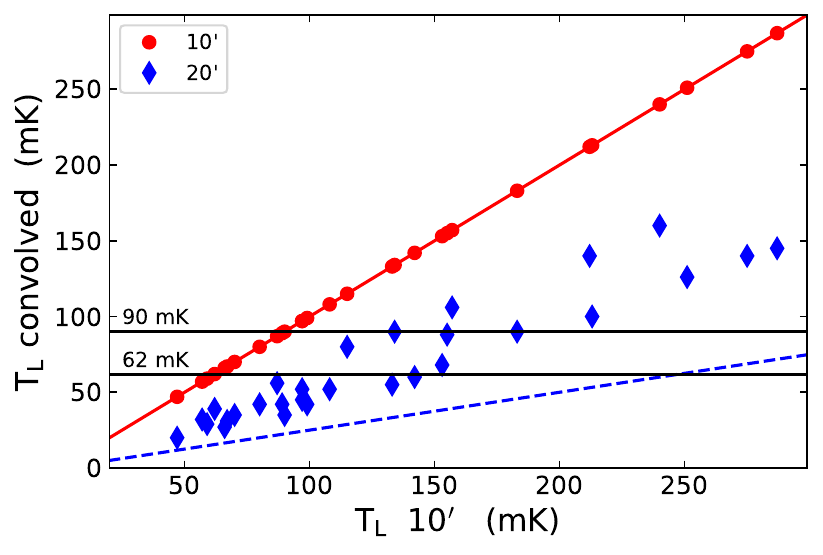}
\caption{
The distance-dependence of line brightness, \TL, for clouds in the deep-near sample, estimated from an average of the data over a $20'\times 20'$  area.
This shows the faintest $75\%$ of the clouds, but the brighter clouds fall in the same pattern.
Each cloud appears twice: first as measured over the basic $10'\times10'$ area used throughout this work  (red circles), then for an average over a $20'\times 20'$ rectangle giving the \TL\  as it would be derived were it at twice its actual distance (blue diamonds).
The measurements are plotted against the \TL\ derived from the $10'$ average.
The inverse-square dependence of $\TL$ on distance expected for very small clouds is shown by the blue dashed line for a factor of two distance change.
All the clouds lie well above the inverse-square line, indicating that the  extended \hi\ structure of the clouds results in a distance-dependence that is shallower than inverse-square.
The two horizontal lines show  the GBT survey $3\sigma$ detection limits -- the upper line for the general survey (labeled ``90 mK'') and the lower line for the 100 sq-deg of the deep survey area (labeled ``62 mK'').
Clouds that lie below these lines (blue diamonds) would not reach the $3\sigma$ noise limit of the survey were they on the far side of the FB -- most would not have been detected.
Note that these are empirical results and do not depend on knowledge of the distance to any cloud.
}
\label{fig:TL_scaled}
\end{figure}

\end{document}